\documentclass[acmlarge]{acmart}

\usepackage{multirow}
\usepackage{subcaption}
\usepackage{color}

\AtBeginDocument{%
  }

\acmJournal{TOIS}
\acmYear{202x} 
\acmVolume{To appear} 
\acmNumber{To appear} 
\acmArticle{} 
\acmMonth{1} 
\acmDOI{10.1145/3568394}

\setcopyright{none}

\makeatletter
\def\runningfoot{\def\@runningfoot{To appear at}}
\def\firstfoot{\def\@firstfoot{To appear at}}
\makeatother

\begin{document}

\title{The Power of Selecting Key Blocks with Local Pre-ranking for Long Document Information Retrieval}


\author{Minghan Li}
\email{Minghan.Li@univ-grenoble-alpes.fr}
\affiliation{%
  \institution{Université Grenoble Alpes}
  \city{Grenoble}
  \country{France}
}

\author{Diana Nicoleta Popa}
\authornote{The work was done while the author was at Université Grenoble Alpes.}
\email{diana.nicoleta.popa@telepathy.ai}
\affiliation{%
  \institution{Telepathy Labs}
  \city{Zurich}
  \country{Switzerland}
}

\author{Johan Chagnon}
\authornotemark[1]
\email{jjrc826@uowmail.edu.au}
\affiliation{%
  \institution{University of Wollongong}
  \city{Wollongong (NSW)}
  \country{Australia}
}

\author{Yagmur Gizem Cinar}
\authornotemark[1]
\email{cinary@amazon.com}
\affiliation{%
  \institution{Amazon}
  \city{Edinburgh}
  \country{United Kingdom}
}

\author{Eric Gaussier}
\email{Eric.Gaussier@univ-grenoble-alpes.fr}
\affiliation{%
  \institution{Université Grenoble Alpes}
  \city{Grenoble}
  \country{France}
}

{\let\thefootnote\relax\footnotetext{An earlier and short version of this article was presented at SIGIR 21, titled: ``KeyBLD: Selecting Key Blocks with Local Pre-ranking for Long Document Information Retrieval'' \cite{li2021keybld}.}}
\renewcommand{\shortauthors}{Li et al.}

\begin{abstract}
  On a wide range of natural language processing and information retrieval tasks, transformer-based models, particularly pre-trained language models like BERT, have demonstrated tremendous effectiveness. 
  Due to the quadratic complexity of the self-attention mechanism, however, such models have difficulties processing long documents. Recent works dealing with this issue include truncating long documents, in which case one loses potential relevant information, segmenting them into several passages, which may lead to miss some information and high computational complexity when the number of passages is large, or modifying the self-attention mechanism to make it sparser as in sparse-attention models, at the risk again of missing some information. 
 We follow here a slightly different approach in which one first selects key blocks of a long document by local query-block pre-ranking, and then few blocks are aggregated to form a short document that can be processed by a model such as BERT. Experiments conducted on standard Information Retrieval datasets demonstrate the effectiveness of the proposed approach.
\end{abstract}



\maketitle

\section{Introduction}

The field of query-document information retrieval (IR) has seen increasingly rapid advance in the past decades. Learning-to-rank (LTR) models \cite{Liu2011,li2011learning} have already achieved great success in many IR applications. However, LTR models mainly rely on hand-crafted features which are time-consuming and often over-specific in definition \cite{guo2020deep}. With the resurgence of interest in neural networks, especially deep neural networks or deep learning, we have witnessed dramatic improvements in computer vision and natural language processing (NLP) tasks. Neural Information Retrieval (Neural IR), which refers to the use of (deep) neural networks to directly construct the
ranking function for IR, has been the subject of many studies \cite{huang2013learning,shen2014latent,palangi2016deep,guo2016deep,pang2016text,pang2017deeprank,hui2017pacrr,xiong2017end,fan2018modeling,li2018nprf} which have led to the development of several interesting IR models for learning representations of documents and queries.

The transformer model \cite{vaswani2017attention}, which is exclusively based on multi-head attention mechanism, has shown to be higher in quality while being more parallelizable and requiring substantially less training time than models based on recurrent neural networks
\cite{vaswani2017attention}.
Using a multi-layer bidirectional transformer encoder, the authors of \cite{devlin2018} have proposed Bidirectional Encoder Representations from Transformers (BERT), a method for pre-training deep bidirectional representations from
unlabeled text by conditioning all layers on both left and right context. Pre-trained BERT models can be fine-tuned to produce cutting-edge models for a variety of applications.
In particular, following their success in NLP, several works have focused on transformers \cite{vaswani2017attention} and derived models based on BERT \cite{devlin2018} for solving IR tasks \cite{nogueira2019passage,macavaney2019cedr,dai2019deeper,li2020parade}, leading to some of the current state-of-the-art models in \textit{ad hoc} IR \cite{macavaney2019cedr,li2020parade}. 

One main advantages of the self-attention mechanism is that it allows to capture dependencies between tokens in a sequence regardless of their distance. 
However, despite its excellent results, the self-attention mechanism has difficulty to process long sequences due to its quadratic complexity in the number of tokens, which also limits the application of transformer-based models to long document information retrieval, where each document could contain thousands of tokens.

Three standard strategies based on BERT have been adopted to circumvent this problem. The first one consists in truncating long documents (\textit{e.g.}, \cite{nogueira2019passage}), the second in segmenting long documents into shorter passages (e.g., \cite{dai2019deeper}) and the last one in replacing the complex self-attention module with sparse-attention ones (e.g., \cite{jiang2020long}). In the first case, important information may be lost and the relevance judgement is damaged. In the second case, a hierarchical architecture can be further adopted to build a document-level representation on top of the representations of each passage \cite{li2020parade}. This said, despite the state-of-the-art results this strategy may lead to, there remain issues concerning the time, memory and energy consumption associated to it. Furthermore, the consideration of passages that may not be relevant to the query may introduce noise in the final representation and limit the identification of long-distance dependencies between relevant tokens \cite{NEURIPS2020_96671501}. In the third case, sparsity constraints may lead to miss important dependencies which can lead to under-optimal results. 
 
The approach we propose is slightly different with these strategies, aiming at capturing, in long documents, the blocks which are the most important to decide on the relevance status of the whole document. Besides, it can be integrated to the second strategy. It is based on three main steps: (a) selecting key (i.e., likely relevant) blocks with local pre-ranking using either classical IR models or a learning module reminiscent of the judge module used in \cite{NEURIPS2020_96671501}, (b) learning a joint representation of queries and key blocks using a standard BERT model, and (c) computing a final relevance score which can be regarded as an aggregation of local relevance information \cite{wu2007retrospective}.

Our contributions are two-fold. We first conduct an analysis which reveals that relevance signals can appear at different locations in documents and that such signals can be better captured by semantic relations than by exact matches. We then investigate two methods to select blocks, one based on standard IR functions and the other on a learned function operating on semantic representations, and show how to integrate these methods in state-of-the-art IR models. In this approach, as well as in previous approaches based on passages as PARADE \citep{li2020parade}, blocks occurring at different positions of a document are concatenated or selected in the order they occur in the document and can be seen as a digest of the elements necessary to assess the relevance of the whole document to the query. Although the blocks selected are not coherent physically, they still are coherent in that they are all relevant to the same query.

The remainder of the paper is organized as follows: Section~\ref{sec:related-work} describes related work. Section~\ref{sec:analysis} investigates the relation between the potential relevance and the position of a block in a document as well as the importance of fuzzy vs exact matching when selecting blocks. Section~\ref{blockselectSec} presents the block selection approach based on standard IR functions whereas Section~\ref{keyb3Sec} describes the block selection approach based on a learned function. Sections~\ref{sec:experiments} and \ref{sec:experiments-2} show the benefits of selecting key blocks on different collections. Finally, Section~\ref{sec:conclusion} summarizes our findings and concludes the paper.

\section{Related work}
\label{sec:related-work}

Let $q$ denote a query and $d$ a document. Without loss of generality, the ranking function $f$ of an IR system takes the form \cite{guo2020deep}:
\begin{equation}\label{Eq.unified_formulation}
	f(q, d) = g(\psi(q), \phi(d), \eta(q,d)),
\end{equation}
where $\psi$ and $\phi$ are representation functions that extract features from $q$ and $d$ respectively, $\eta$ is the interaction function that models query-document representation from $(q,d)$ pairs, and $g$ is the evaluation function that calculates the relevance score based on the extracted features or interaction. According to the choices made on the representation and interaction functions, neural information retrieval models can be grouped into two categories \cite{guo2020deep}: representation-based and interaction-based architectures. Besides these two categories, some neural information retrieval models adopt a hybrid approach.

The Deep Structured Semantic Model (DSSM) \cite{huang2013learning} is one of the earliest representation-based models for document ranking which uses a fully-connected network for the functions $\psi$ and $\phi$. To map the query and the documents to a shared semantic space, a non-linear projection is used. The relevance of each document given the query is then calculated with the cosine similarity between their vectors in that semantic space. Clickthrough data is then used to discriminatively train the model by maximizing  the conditional likelihood of the clicked documents. Other studies in this category proposed to exploit distributed representations via DSSM variations, or relied on different representation functions \cite{onal2018neural}. For example, ARC-I \cite{hu2014convolutional} and CLSM \cite{shen2014latent} use convolutional neural networks (CNN) for $\psi$ and $\phi$ while \cite{palangi2016deep} uses a recurrent neural networks. 

One of the first neural IR models which outperformed traditional IR models is the interaction-based model referred to as Deep Relevance Matching Model (DRMM) \cite{guo2016deep}. The interaction function $\eta$ is defined as the matching histogram mapping between each query term and the document. A feed-forward network for term-level relevance and a gating network for score aggregation in the evaluation function $g$ are further used. In this work, the term vectors are fixed to Word2Vec word embeddings \cite{NIPS2013_9aa42b31}. Similarly, \citet{xiong2017end} proposed KNRM which employs a translation matrix that utilizes word embeddings to represent word-level similarities, a unique kernel-pooling technique for extracting multi-level soft match features, and a learning-to-rank layer that combines those features into the final ranking score. The entire model is trained end-to-end, and the word embeddings are tuned to produce the desired soft matches \cite{xiong2017end}. Inspired by the way humans assess the relevance of a document, \citet{pang2017deeprank} proposed DeepRank, a model  which splits documents into term-centric contexts according to each query term. A tensor containing both the word representations of query/query-centric context as well as their interactions is first built. It is then passed through a measure network, based on CNN \cite{krizhevsky2012imagenet} or 2D-GRU \cite{DBLP:conf/ijcai/WanLXGPC16}, to produce a representation of local relevance. Finally, the global relevance score is calculated using an aggregation network. \citet{hui2017pacrr} proposed PACRR, a model inspired by the neural models used in image recognition \cite{guo2020deep}. PACRR takes a similarity matrix between a query and a document as input. Then multiple CNN kernels capture the query-document interactions. Following this work, \citet{hui2018co} provided a lightweight contextualization model called CO-PACRR which averages word vectors within a sliding window and appends the similarities to those of the PACRR \cite{hui2017pacrr} model \cite{DBLP:conf/ecai/HofstatterZH20}.

Since it is sometimes difficult to produce good high-level representations of long texts, the representation-based architecture is better suited to short input texts. Models in this category are good for online computing since they allow one to pre-calculate text representations. Interaction-based models, on the other hand, tend to yield better results as they can tune document representations towards a given query. Unfortunately, since the interaction function $\eta$ cannot be pre-calculated until the input pair $(q, d)$ is seen, models in this category are not as efficient for online computation as representation-focused models \cite{guo2020deep}.

\paragraph{Models Based on transformers}
Benefiting from pre-trained language models based on transformers \cite{vaswani2017attention}, especially BERT \cite{devlin2018}, different research teams have developed state-of-the-art neural IR models, significantly outperforming traditional and previous neural IR models. \citet{nogueira2019passage} proposed to use BERT as a re-ranker for the passage re-ranking task by fine-tuning it and achieved state-of-the-art results. The passages are truncated if too long (typically over 512 tokens). This work proved the effectiveness of fine-tuning BERT for IR problems. \citet{macavaney2019cedr}, through a model called CEDR which combined BERT with other neural IR models, as PACRR \cite{hui2017pacrr}, KNRM \cite{xiong2017end} and DRMM \cite{guo2016deep}, and showed the benefits of this combination. 
\citet{dai2019deeper} proposed to first segment documents into short, overlapping passages, and then used BERT to define the relevance score of the document, using either the first passage, the best passage or the sum of all passages. \citet{DBLP:conf/ecai/HofstatterZH20} proposed a reranking model called Transformer Kernel, in short TK, which uses a hybrid approach based on a small number of transformer layers to contextualize query and document word embeddings separately. Then RBF-kernels \cite{xiong2017end} are used for interaction scoring, where each kernel focuses on a specific similarity range. Experimental results show that although the effectiveness is not as good as BERT reranker, TK has strong efficiency. In a similar vein, \citet{li2020parade} explored strategies for aggregating relevance signals from a document’s passages into a final ranking score, leading to a model called PARADE. A hierarchical layer, in the form of a max-pooling, attention, CNN or transformer aggregator is used to aggregate the passage representations so as to obtain a joint query-document representation for long documents. They showed that passage representation aggregation strategies can outperform techniques proposed previously. In particular, PARADE can improve results significantly on collections with broad information needs where relevance signals can be disseminated throughout the document. \citet{Grail2021} also proposed a hierarchical model in which each transformer layer, used to learn a representation for each sentence of a document, is followed by an RNN which captures dependencies between the CLS tokens representing the different sentences of a document. As shown in the experiments, this model is particularly well adapted for long-document summarization. 

In the above models, as transformers are limited in their input length due to their quadratic complexity, researchers have either truncated long documents or segmented them into passages. There have however been different attempts to use transformers on long documents. For example, \citet{dai2019xl} introduced a model with left-to-right recurrence between transformer windows, consisting of a segment-level recurrence mechanism and a novel positional encoding scheme. The left-to-right approach processes the document in chunks moving from left-to-right and thus not adapted to tasks which benefit from bidirectional contexts \cite{beltagy2020longformer}. \citet{child2019} introduced several sparse factorizations of the attention matrix which reduce the quadratic complexity to $O(n\sqrt{n})$. \citet{hofstatter2020local} proposed a local self-attention which considers a sliding window over the document and restricts the attention to that window in order to deal with long documents. Their model, called TKL, adapts TK \cite{DBLP:conf/ecai/HofstatterZH20} with this mechanism.
\citet{beltagy2020longformer} introduced the Longformer with an attention mechanism which scales linearly with sequence length, combining windowed local-context self-attention with task-motivated global attention to encode inductive bias about the task. Longformer achieves state-of-the-art results on the character-level language modeling tasks, and when pretrained from the RoBERTa \cite{liu2019roberta} checkpoint, it consistently outperforms RoBERTa on long document tasks. \citet{Zhao2020Transformer-XH:} proposed Transformer-XH which enables to represent structured texts. It shares similar motivation with \citet{dai2019xl} and \citet{child2019}, and is particularly well adapted to multi-hop QA tasks \cite{yang2018hotpotqa} and fact verification tasks \cite{thorne2018fact}. \citet{ainslie2020etc} introduced the Extended Transformer Construction (ETC) model to address two key challenges of standard transformers: scaling input length and encoding structured inputs. A novel global-local attention mechanism is introduced where the local sparsity reduces the quadratic scaling of the attention mechanism. 
They further show that by including a pre-training Contrastive Predictive Coding (CPC) task \cite{oord2018representation}, the performance for tasks where structure matters improves even further. \citet{zaheer2021big} proposeed BigBird, which combines local and global attention with random sparse attention. \citet{kitaev2020reformer} only computed self-attention between similar tokens, as defined through locality-sensitive hashing.

Despite such models' described effectiveness, there remain problems. Firstly, as described in \cite{zaheer2021big}, coalesced memory operations, which load blocks of contiguous bytes at once, are where hardware accelerators like GPUs and TPUs really shine. As a result, small sporadic look-ups caused by a sliding window or random element queries are not very efficient. This is addressed by "blockifying" the lookups. It is generally known \cite{gray2017gpu,yao2019balanced} that GPUs cannot efficiently execute sparse multiplications, which are commonly employed by models with tailored attention mechanisms. Naively using for-loops or masking the matrix may result in even worse efficiency than the full self attention \cite{jiang2020long}. 
Thus, such models with customized attention mechanisms need specifically designed tricks or customized CUDA kernels \cite{beltagy2020longformer}, which are inconvenient or require expertise in low-level GPU operations \cite{jiang2020long}. \citet{jiang2020long} proposed Query-Directed Sparse Transformer (QDS-Transformer), which also induce sparsity in self-attention mechanism, containing local windowed attention and query-directed
global attention. Experiments demonstrate consistent and robust advantages of QDS-Transformer over previous approaches. However, this model still needs customized TVM implementation \cite{chen2018tvm}. Secondly, as these customized attention models only rely on partial attention, their accuracy does not match, in general, the one of full attention models.

\paragraph{Selecting key blocks} The approach we advocate here to solve the above-mentioned problems is to select key, important blocks from a document and base the relevant score of the full document on just these blocks. Note that this differs from the approach followed in PARADE \cite{li2020parade} in which the passages retained are arbitrary. It is however reminiscent of both \cite{pang2017deeprank} and \cite{NEURIPS2020_96671501}, which are partly inspired by cognitive theory and reckon that, in order to assess the relevance of a document, humans first scan the whole document to detect relevant locations where query terms occur and then aggregate local relevance information to decide on the relevance of a document \cite{wu2007retrospective}. Compared to \cite{pang2017deeprank}, our approach is simpler conceptually and can consider blocks which do not contain query words but are nevertheless relevant to the query. We show in Sections~\ref{sec:analysis} and \ref{sec:experiments} that such a \textit{fuzzy matching} can help improve the block selection process and the overall IR system built upon it. The study described in \cite{NEURIPS2020_96671501} focuses on reading comprehension, multi-hop question-answering and text classification. We show here how this approach can be simplified for IR purposes by using the same BERT model for the \textit{reasoner} and \textit{judge}. In addition, we investigate the use of standard models to select key blocks, which leads to an entirely different and simpler architecture.

Furthermore, we want to mention the study presented in \citet{muntean2020weighting} which, in order to assign a relevance score to a document, weighs each passage differently by identifying salient terms using TF-IDF and KL divergence scores \cite{buttcher2006document} are used to identify salient terms and to derive the weights. Although this model treats passages in a long document differently, the weights do not reflect the relevance to the query (salient terms are identified independently of the query). This is a major difference with our approach which aims to select blocks according to their relevance to the query.  Furthermore, we focus here on neural IR models which have difficulties in dealing with long documents.

This paper is an extension of the earlier short paper published at SIGIR 2021 and entitled: ``KeyBLD: Selecting Key Blocks with Local Pre-ranking for Long Document Information Retrieval'' \cite{li2021keybld}. This extension consists first in an analysis of where relevant signals appear in documents and of how blocks should be selected, and second in the proposal of a new learning-based selection method (see Section \ref{keyb3Sec}) and an integration within PARADE  \cite{li2020parade} (see Section \ref{improveParadeSec}). We have also added three new collections and several baseline models for evaluation purposes. Experimental results show the effectiveness of the proposed approaches for selecting key blocks. In particular, the method we developed for learning how to select blocks outperforms all other methods, including the ones we presented in \cite{li2021keybld}. Besides, selecting key passages with local pre-ranking makes the PARADE model more efficient and accurate: the results obtained are in general better or at least similar. This shows that the proposed mechanism can be deployed in different models. In addition, our approach leads to better results than sparse attention transformer models while not requiring customized CUDA kernels.

Lastly, in parallel to our work, \citet{DBLP:conf/sigir/HofstatterMZCH21} also introduce a model called IDCM which also learns to select top scoring blocks which are then used to score documents with respect to a given query. IDCM first trains a block ranking model based on BERT (and called ETM for Effective Teacher Model) on MS-MARCO\footnote{MS-MARCO is a passage retrieval collection available at: \url{https://microsoft.github.io/msmarco/}}, prior to fine-tuning this model on each collection for document ranking. Then, a block selection model  (called ESM for Efficient Student Model) is learned via knowledge distillation from ETM. Both ESM and ETM are then used to score new documents, ESM allowing the selection of the most important blocks and ETM being used to score documents on the basis of the selected blocks. This contrasts with the way we learn to select blocks: first of all, for all proposed models except the last one KeyB(PARADE5)$_{BinB2}$,  we do not require any pre-training on additional collections; second, we use the same model for selecting blocks and scoring documents, the rationale being that in both cases one computes a relevance score with respect to the same query. The last model that we propose can also re-use an additional BERT ranker from the KeyB(vBERT) models previously proposed to select blocks.  Our approach is thus simpler (see Section~\ref{keyb3Sec}) and finally leads to better results as shown in Section~\ref{sec:experiments-2}. Furthermore, it is interesting to note that IDCM scores each block separately and then aggregates the block  scores, while our approach for improving Vanilla BERT scores a document using concatenated selected blocks. As shown in our experiments, the two strategies are effective, with a slight advantage for the latter on the TREC 2019 DL collection. We also want to mention that we recently proposed \citet{li2022bert} to select key blocks for late-interaction models by training a BERT-based model with multi-task learning. This study however has a different focus that the current one, which only concerns interaction-based models.

\section{A finer-grained look at documents}
\label{sec:analysis}

We take in this section a closer look at documents and the blocks they contain by addressing two questions:
\begin{enumerate}
\item Are relevant signals spread over the entire document, and thus can appear in any block, or are they concentrated in particular regions, as the beginning of documents?
\item Should one rely only on exact matching of query words to select important blocks or is additional information contained in related words, as synonyms, important?
\end{enumerate}
Our analyses thus aim to reveal which blocks to select with respect to their positions and how to select them.

\subsection{Preliminaries}

To conduct our investigation of the above points, we use four standard IR datasets, namely MQ2007, MQ2008, GOV2\footnote{\url{http://ir.dcs.gla.ac.uk/test_collections/gov2-summary.htm}} (also referred to as Trec-terabyte 2004/2005/2006) and Robust04. MQ2007 and MQ2008 are standard LETOR \cite{qin2013introducing} benchmark datasets for learning to rank. GOV2 contains documents resulting from a crawl of .gov websites made in early 2004. Robust04\footnote{\url{https://trec.nist.gov/data/robust/04.guidelines.html}} contains news article from the Financial Times, the Federal Register 94, the LA Times, and FBIS. 
In each dataset, the title of the topics have been used as queries and the content of the documents have been extracted using Anserini \cite{yang2018anserini}. Relevance judgements can take three values: 0 (irrelevant), 1 (relevant) or 2 (very relevant). A document-query pair with an associated relevant judgement will be referred to as a \textit{labeled} document-query pair. All our analyses are based on labeled document-query pairs so as to avoid assumptions on the relevance status of non-labeled pairs. Furthermore, as it has become common to first filter documents with a standard IR system prior to deploy deep IR models, we first select, for each query, the top 200 documents using BM25 and only retain the labeled document-query pairs associated with these documents. Note that the above filtering is not run on MQ2007 and MQ2008 which already rely on a small subset of documents for each query. Table \ref{tab:datasets} displays, for each collection, the number of queries and documents as well as the number of unique labeled document-query pairs (in the original dataset as well as the one filtered with BM25). The proportions of irrelevant, relevant and very relevant pairs are computed on the filtered version for GOV2 and Robust04, and on the original version for MQ2007 and MQ2008.

\begin{table}[ht]
    \centering
    \begin{tabular}{lrrrr}
    \toprule
        Dataset                     & MQ2007 & MQ2008 & GOV2    & Robust04 \\ \midrule
        Nb of queries               & 1,692  & 784    & 150     & 250      \\
        \multicolumn{5}{l}{\textbf{Original}} \\ 
        Nb of documents   & 65,302 & 14,381 & ca. 25M & ca. 0.5M  \\
        Nb of labeled document-query pairs             & 69,599 & 15,208 & 135,352 & 311,410  \\
         \multicolumn{5}{l}{\textbf{BM25 filtered}} \\
        Nb of unique documents  & - & -   & 29,769  & 42,156  \\
        Nb of labeled document-query pairs    & - & - & 26,155 & 95,336  \\ \hline
        Proportion of irrelevant pairs   & 0.74   & 0.81   & 0.80    & 0.94     \\
        Proportion of relevant pairs     & 0.20   & 0.13   & 0.17    & 0.05     \\
        Proportion of very relevant pair & 0.06   & 0.6   & 0.03    & 0.01    \\
    \bottomrule
    \end{tabular}
    \caption{Statistics of the datasets used.}
    \label{tab:datasets}
\end{table}

To divide documents into blocks, we use the recent CogLTX \cite{NEURIPS2020_96671501} block decomposition method, which is based on a dynamic programming method, each block having a maximum of 63 tokens. This method, which was used with success on several NLP tasks, sets different costs for different punctuation marks and aims at segmenting in priority on strong punctuation marks such as "." and "!", making it close to a sentence segmentation procedure.

Lastly, to measure to which extent a block is relevant or not to a given query, we use the Retrieval Status Value (RSV) of the block which simply amounts here to the score provided by an IR model, in our case either BM25 or the cosine similarity computed on semantic representations of queries and blocks. 
The higher the score obtained, the likelier the corresponding block is relevant. 

We now turn to answer the two questions we asked before.

\subsection{Relevance signals appear at different positions in documents}

\begin{figure}[ht]
    \centering
    \includegraphics[width=\textwidth]{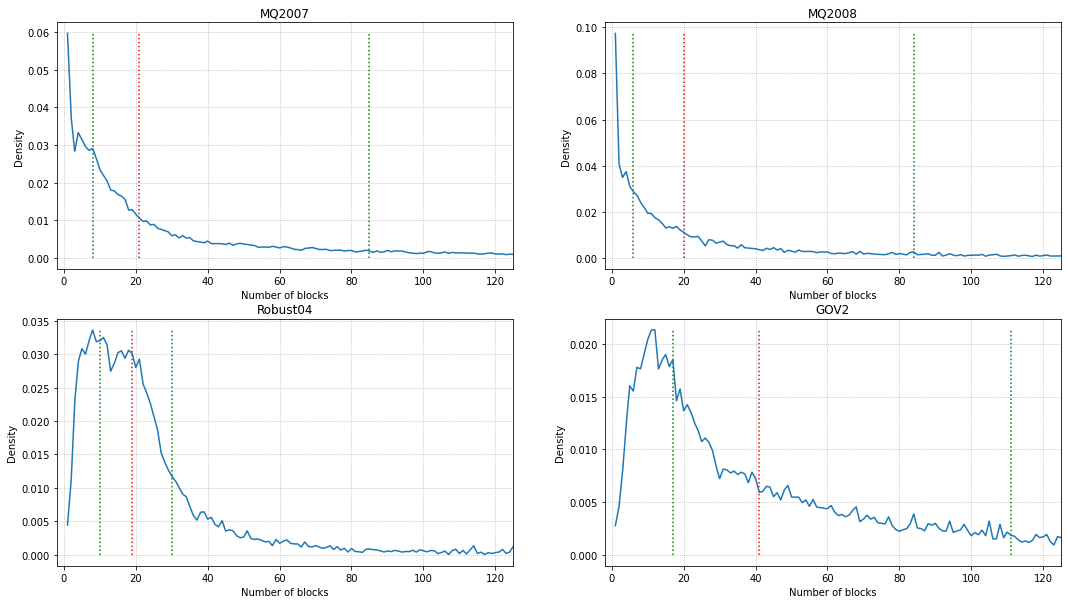}
    \caption{Interpolated curve of the density of the number of blocks per document for the different datasets. The vertical line in red denotes the position of the median, whereas the position of the 1st and 3rd quartile are displayed in green. For the sake of readability, only the first 125 blocks are shown here.}
    \label{fig:doc_len}
\end{figure}

We first analyze the length of documents with respect to the number of blocks they contain. To do so, we plot in Figure~\ref{fig:doc_len} the proportion of documents containing exactly $n$ blocks, $n$ varying from $1$ to more than $125$\footnote{The maximum number of blocks for MQ2007, MQ2008, Robust04 and GOV2 respectively is  3225, 3225, 2959, 3311. We do not display the entire distribution for reading purposes.}. In addition, the median as well as the first and third quartiles of the distribution of the number of blocks for each dataset are provided. Even though the shape of the curve for each dataset varies, one can notice that more than 25\% of the documents in MQ2007, MQ2008 and GOV2 contain more than 80 blocks, way above the size limitation of current transformer-based IR models. For Robust04, 25\% of the documents have more than 30 blocks, a size that can also exceed the limitation of current transformer-based IR models\footnote{The average number of tokens per block for MQ2007, MQ2008, Robust04 and GOV2 respectively is 54.39, 54.44, 53.16, 54.33.}. 

As the number of blocks contained in a document varies a lot from one document to another, for assessing where relevance signals appear in different documents, we consider $p$ positions (${1, 2, \cdots, p}$) and allocate each block of a document to one of the $p$ positions with the constraint that the first block of the document should be allocated to the first position, and the last block to the last position. This can be easily done with the following function which provides the position of the $i^{th}$ block of a document containing $b$ blocks ($1\le i \le b$):
\begin{equation}
  pos(i) =  \left \lceil{ \frac{10\times i}{b}}\right \rceil, \nonumber
\end{equation}
where $\lceil x \rceil$ is the ceiling function which returns the integer greater than or equal to $x$. For example, the fifth block in a document containing 100 blocks is in the first position. Only documents that have at least 15 blocks are considered for analysis, preventing missing positions.

\begin{figure*}[tb]
  \hspace{-0.8in}
  \begin{subfigure}[t]{0.3\textwidth}
      \centering
      \includegraphics[height=1.9in]{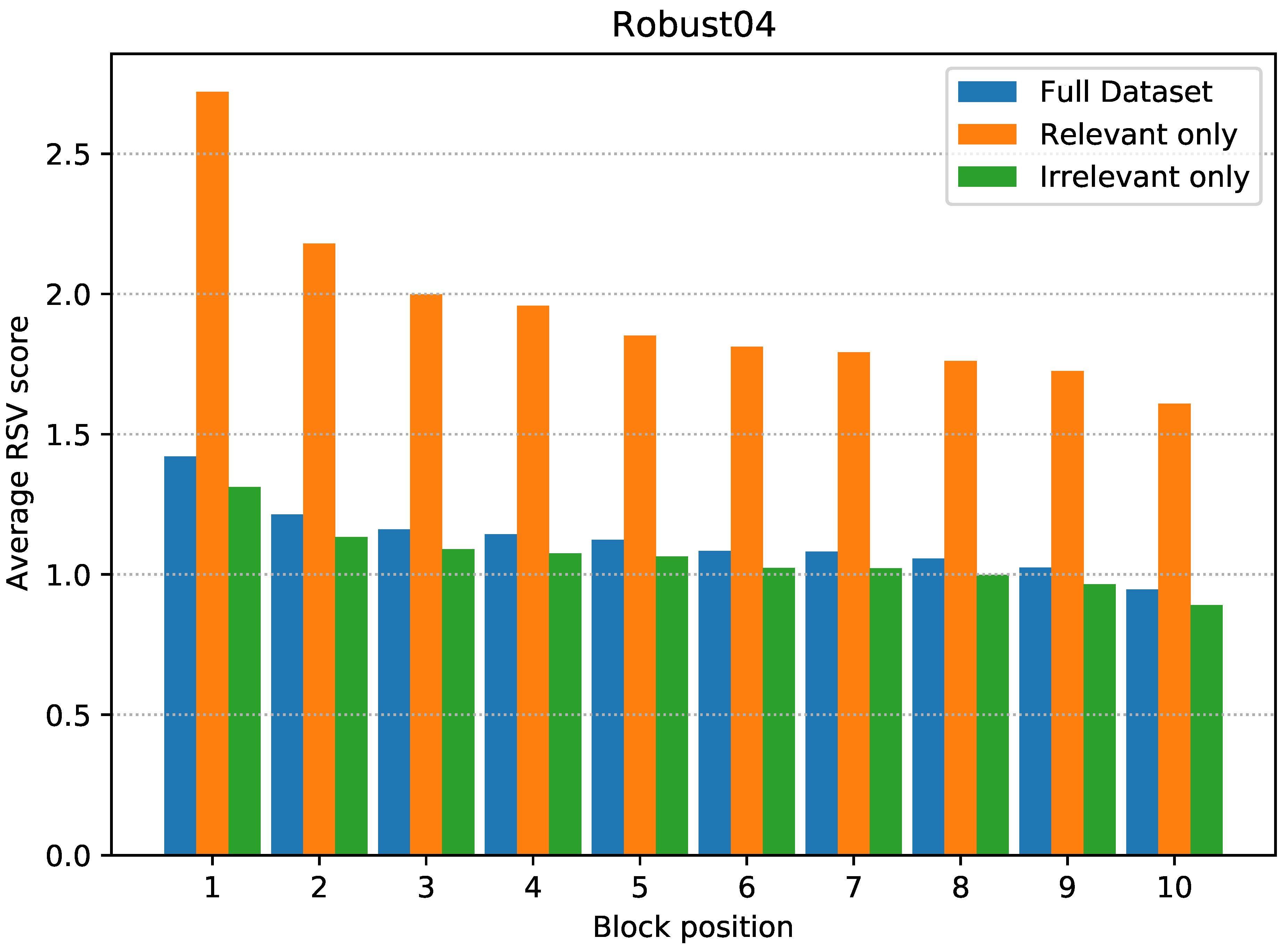}
      \caption{BM25 RSV of Robust04}
  \end{subfigure}%
  \hspace{1in}
  \begin{subfigure}[t]{0.3\textwidth}
    \centering
    \includegraphics[height=1.9in]{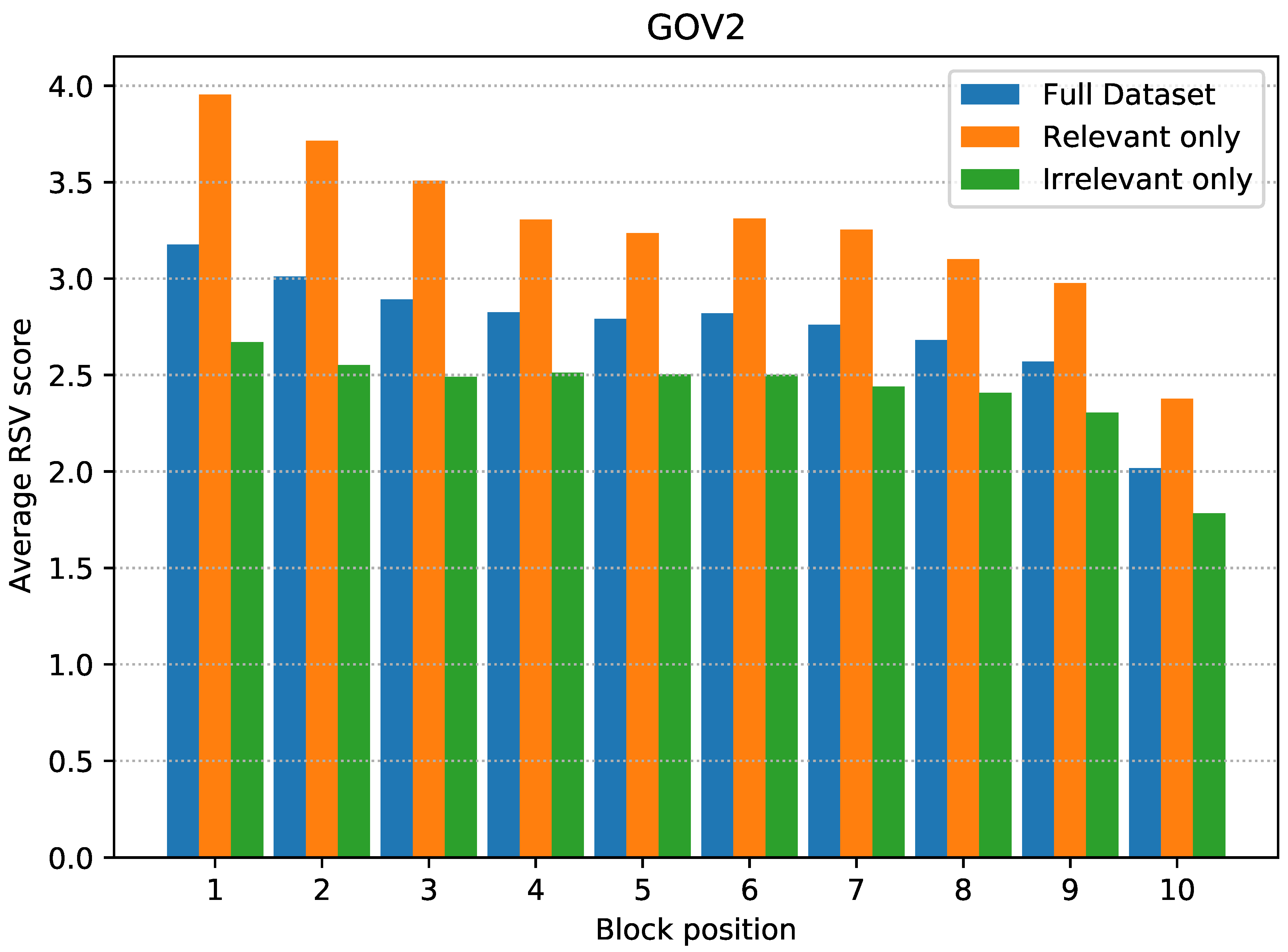}
    \caption{BM25 RSV of GOV2}
\end{subfigure}%

\hspace{-0.8in}
\begin{subfigure}[t]{0.3\textwidth}
  \centering
  \includegraphics[height=1.9in]{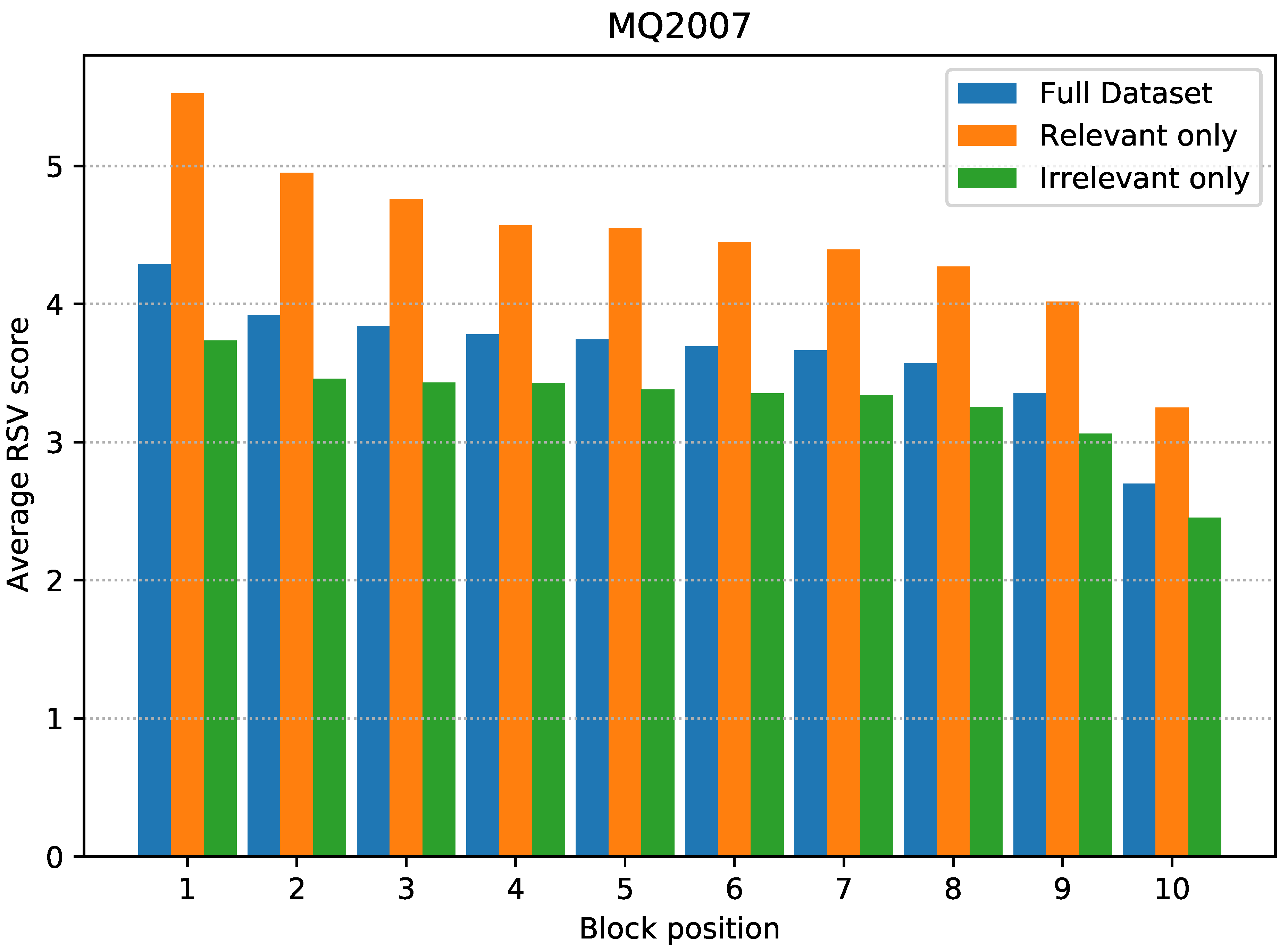}
  \caption{BM25 RSV of MQ2007}
\end{subfigure}%
\hspace{1in}
\begin{subfigure}[t]{0.3\textwidth}
  \centering
  \includegraphics[height=1.9in]{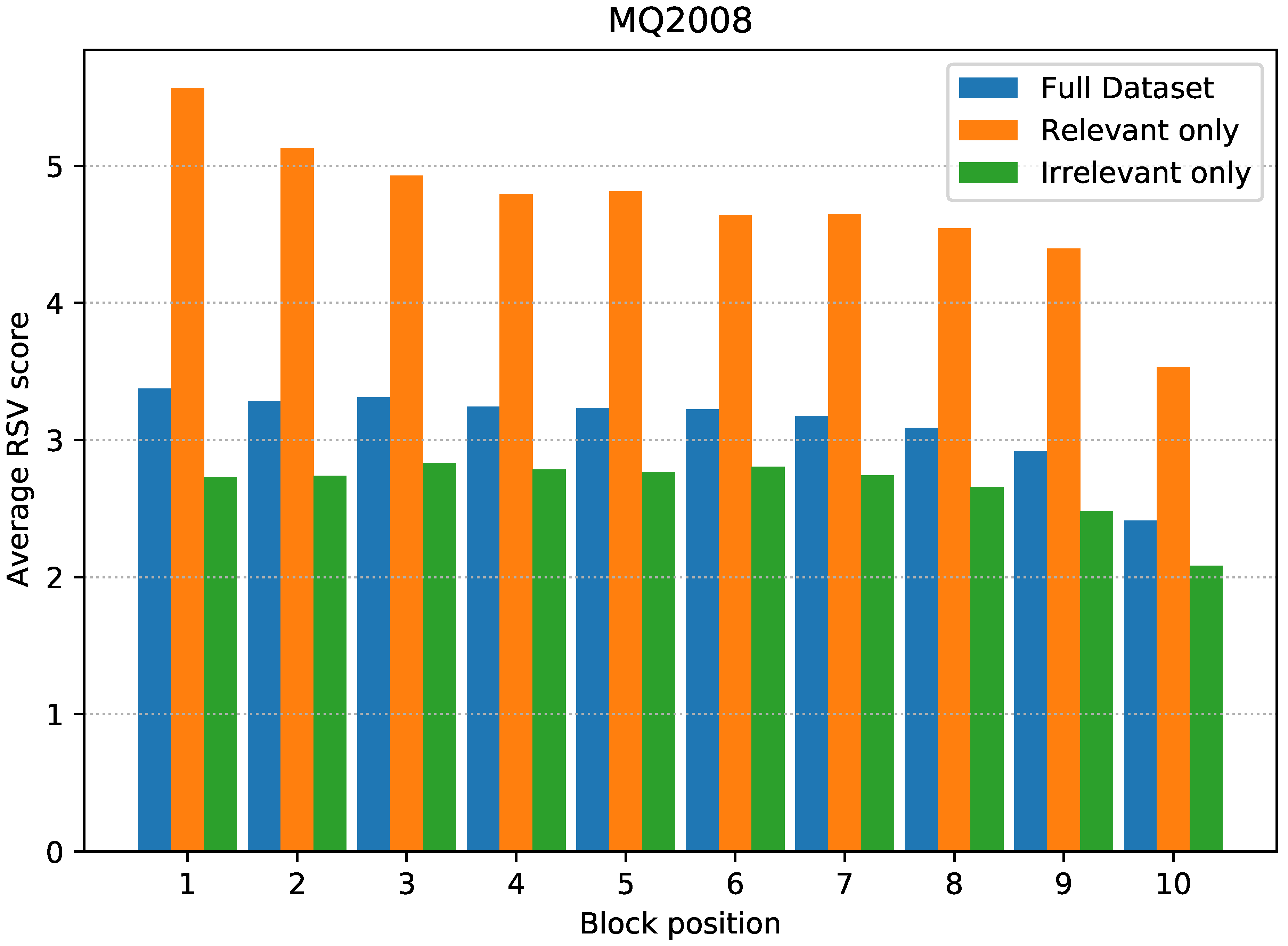}
  \caption{BM25 RSV of MQ2008}
\end{subfigure}%
  \caption{Average BM25 RSV scores per position of a block in a document using the original query $q$.}
  \label{fig:bm25score_bins_mq07_mq08_gov2_r04_q}
\end{figure*}

\begin{figure*}[tb]
  \hspace{-0.8in}
  \begin{subfigure}[t]{0.3\textwidth}
      \centering
      \includegraphics[height=1.9in]{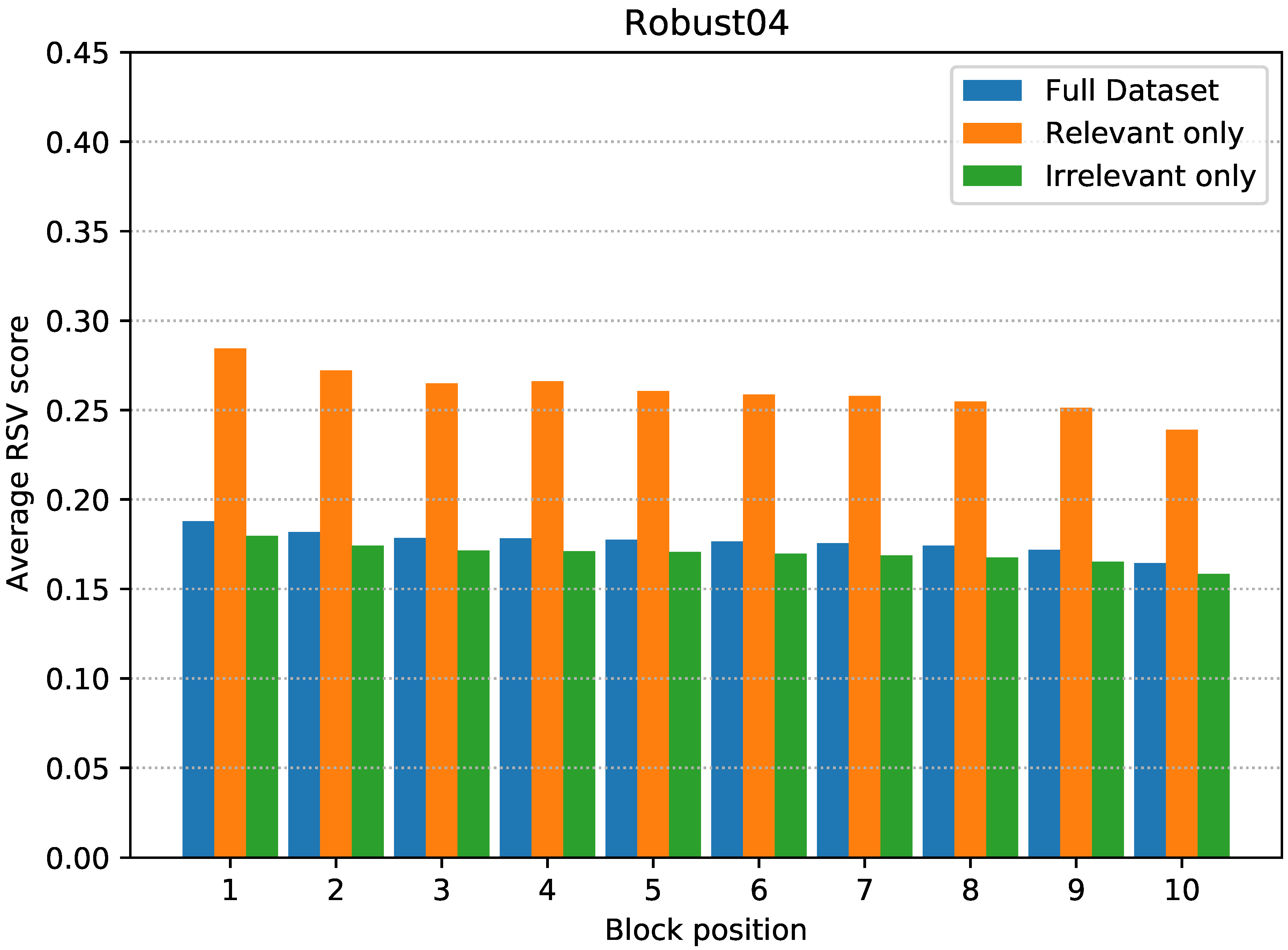}
      \caption{Cosine similarity of Robust04}
  \end{subfigure}%
  \hspace{1in}
  \begin{subfigure}[t]{0.3\textwidth}
    \centering
    \includegraphics[height=1.9in]{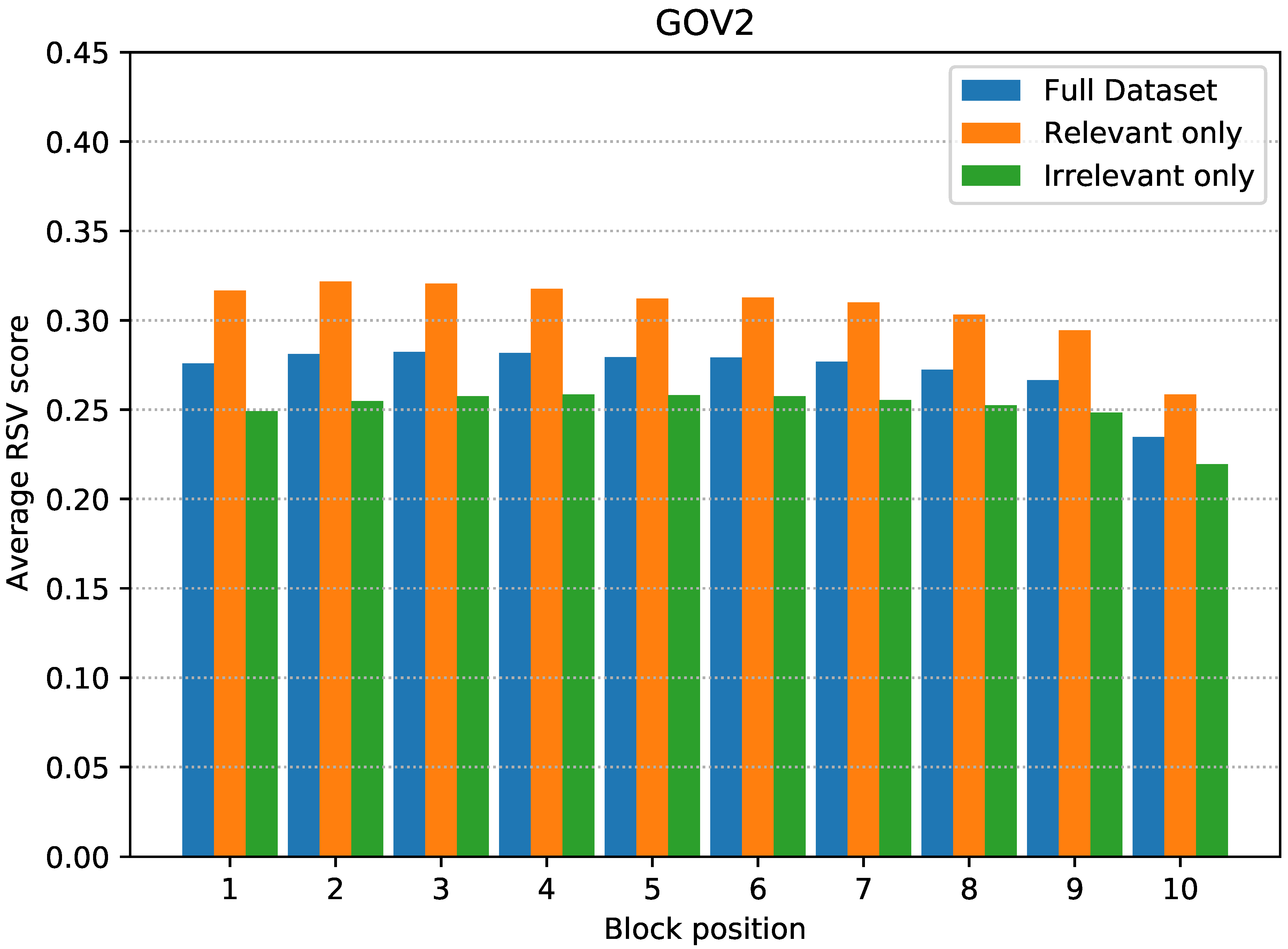}
    \caption{Cosine similarity of GOV2}
\end{subfigure}%

\hspace{-0.8in}
\begin{subfigure}[t]{0.3\textwidth}
  \centering
  \includegraphics[height=1.9in]{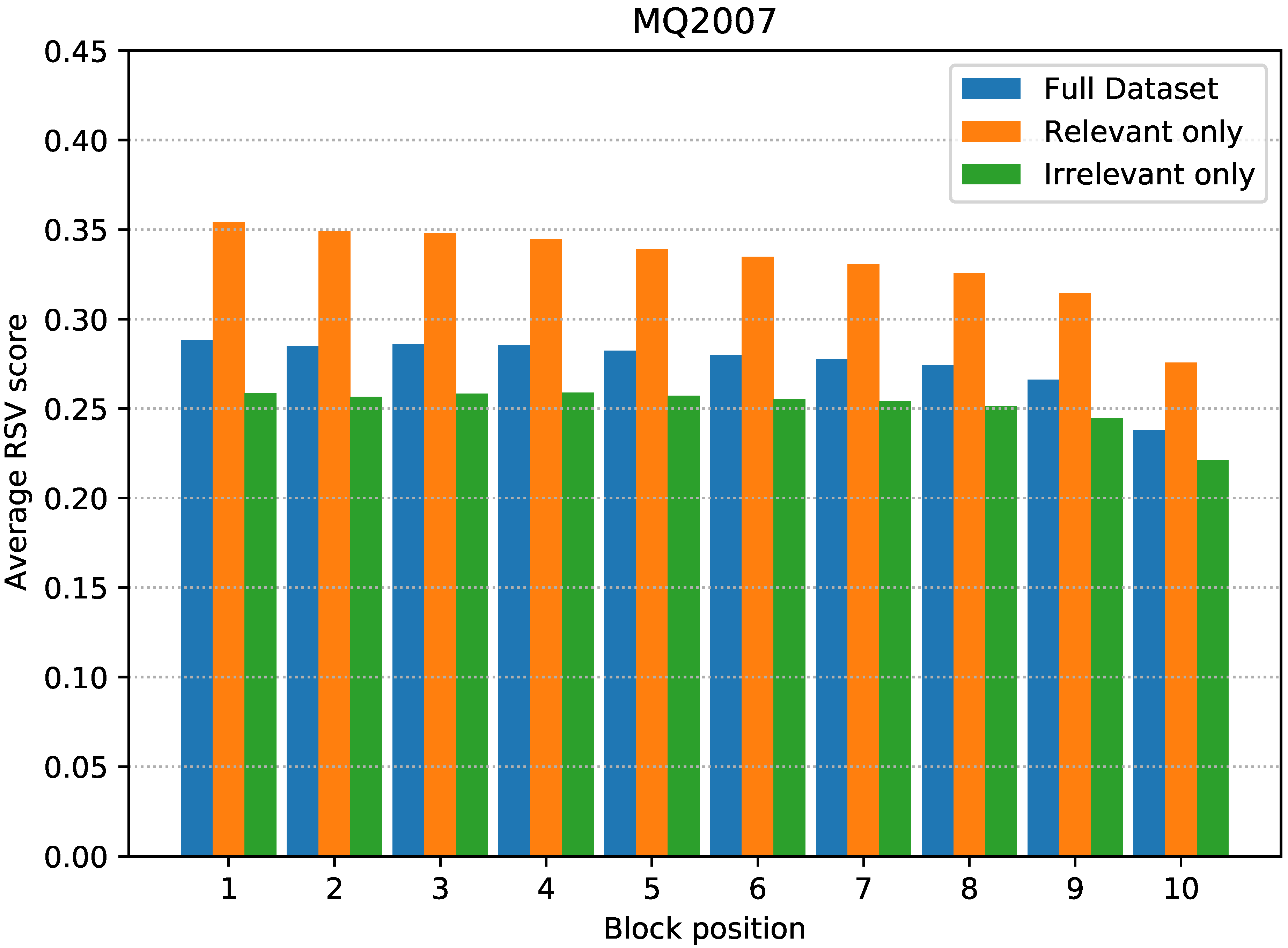}
  \caption{Cosine similarity of MQ2007}
\end{subfigure}%
\hspace{1in}
\begin{subfigure}[t]{0.3\textwidth}
  \centering
  \includegraphics[height=1.9in]{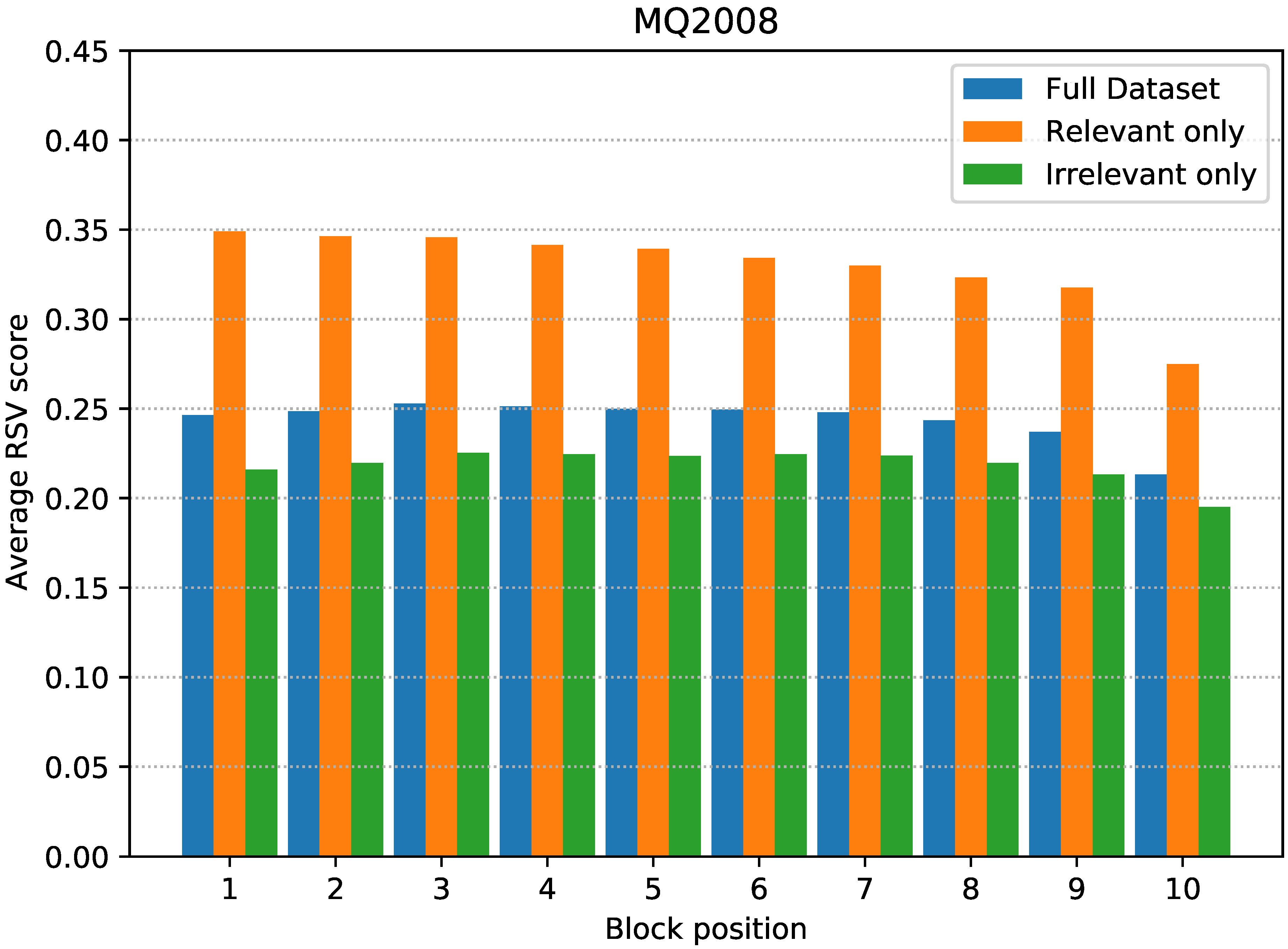}
  \caption{Cosine similarity of MQ2008}
\end{subfigure}%
  \caption{Average RSV scores (cosine similarity) per position of a block in a document using the original query $q$.}
  \label{fig:sbertscore_bins_mq07_mq08_gov2_r04_q}
\end{figure*}

Figure \ref{fig:bm25score_bins_mq07_mq08_gov2_r04_q} displays the distribution of the average BM25 RSV score on each position with $p=10$ for all labeled query-document pairs (referred to as full dataset), for very relevant and relevant pairs (referred to as relevant only) and for irrelevant pairs (referred to as irrelevant only). 
On all collections, the average RSV score decreases when the position increases. However, 
 the average RSV scores are still important in the middle positions and non-negligible in the last positions. 
Furthermore, if one assumes that all blocks of an irrelevant document are irrelevant\footnote{We believe this is a reasonable assumption, at least when the blocks are not too short.}, then the difference, for a given position, between the average RSV scores of relevant and irrelevant pairs may serve as an additional indicator of whether or not relevance signals are found in that particular position. When there is no difference between the average RSV scores of relevant and irrelevant pairs at a given position, then the only conclusion one can draw is that all blocks from relevant and irrelevant documents behave in the same way with respect to the RSV score, and there is no indication that relevance signals are present at that position. 
On the contrary, when the average RSV score for relevant pairs is significantly higher than the one for irrelevant pairs at a given position, then there is a clearer indication that there are relevant signals at that position. This is the case for all positions of the four collections.

To complement the above analysis, we used a different RSV score, namely the cosine similarity between the semantic representations of blocks and queries obtained with the pre-trained Sentence-BERT \cite{reimers2019sentence} model\footnote{The all-MiniLM-L12-v2 version from \url{https://www.sbert.net/docs/pretrained_models.html}} (we simply input each query-block pair to Sentence-BERT which outputs query and block representations on which a cosine similarity is computed). Contrary to the BM25 score which is mainly 'lexical', this score aims to capture additional semantic relationships between queries and blocks. Figure \ref{fig:sbertscore_bins_mq07_mq08_gov2_r04_q} displays the distribution of the average cosine similarity RSV score on each position. This analysis confirms that all positions can contain relevance signals. Furthermore, the decrease in RSV scores when the position of the block increases is less marked so that the difference between blocks at the beginning, the middle and the end of a document (except the last block which gets in general significantly smaller scores) is not really important.

Overall, our analysis reveals that, if terms at the beginning of a document may be more important that terms at the end, all positions in a document can contain important relevance signals and should \textit{a priori} be explored for IR purposes. This conclusion is in agreement with the verbosity hypothesis \cite{Robertson1994}.

\subsection{Fuzzy matching may help select better blocks on some collections}

\begin{table*}[t]
\centering
\caption{Example query and extensions}
\label{tab:example_eldar}
\resizebox{\linewidth}{!}{
\begin{tabular}{ll}
\toprule
\textbf{Original query $q$} & "minimum wage increase" \\ \hline
Synset ("minimum") & minimal \\
Synset ("wage") & earnings, pay, remuneration, salary \\
Synset ("increase") & growth, gain, addition \\ \hline
\multirow{2}{*}{\textbf{Expanded query $q^{exp}$}} & minimum wage increase minimal earnings \\ & pay remuneration salary growth gain addition\\\hline
\multirow{2}{*}{\textbf{Random expanded query $q^{rand\_exp}$}} & minimum wage increase cadent gravely \\ & stuffiness puller complaisant sunlight profusely asterism\\
\hline \hline
Original query boolean representation $q_{bool}$&  ("minimum" OR "wage" OR "increase") \\ \hline
\multirow{4}{*}{Extended query boolean representation $q^{syn}_{bool}$} & (("minimal" OR \\ & "earnings" OR "pay" OR "remuneration" OR "salary" OR \\ & "growth" OR "gain" OR "addition")  \\ & AND NOT ("minimum" OR "wage" OR "increase")) \\\hline
\multirow{4}{*}{Random extended query boolean representation $q^{rand}_{bool}$} & (("cadent" OR \\ & "gravely" OR "stuffiness" OR "puller" OR "complaisant" OR \\ & "sunlight" OR "profusely" or "asterism")  \\ & AND NOT ("minimum" OR "wage" OR "increase")) \\
\bottomrule
\end{tabular}
}
\end{table*}

The second question we address is whether one should solely rely on an exact matching of the words present in the query to select a given block or whether fuzzy matching including words related to the query words may help retrieve better blocks.
We consider here that a word related to a query word is any synonym, as provided by WordNet \cite{wordnet}, of that query word. Other semantic relations can of course be used; we chose synonymy because it has the advantage of being a simple relation which is at least partly captured in modern word embeddings and for which useful resources such as WordNet are available.

Our goal here is to assess whether using synonyms can help select useful blocks. If this is the case, then one can conclude that it may be useful to use matching strategies that go beyond an exact matching of query words. It is important to note here that if many studies have been devoted to the utility of synonyms in IR systems \cite{turney2001mining,li2019multi}, our study differs from them in that it focuses on the use of synonyms for selecting blocks and does not aim to assess different query expansion strategies. In particular, we are not interested in assigning different weights to expanded terms even though such a strategy may lead to better query expansion results \cite{fang2006semantic,fang2008re}.

We first try to answer the question: \textit{Can synonyms identify blocks that would not have been identified without them?} To do so, from the original query, we construct three boolean queries. The first one consists of the disjunction of all query words and will be referred to as $q_{bool}$. The second one consists of the disjunction of all the synonyms from WordNet of all query words and excludes the original query words. It will be referred to as $q^{syn}_{bool}$. Lastly, the third one, referred to as $q^{rand}_{bool}$, which has the same length as the second one, consists of the disjunction of words randomly selected from WordNet. This last query helps assess to which extent the phenomena observed depend solely on the query length\footnote{Adding terms to a boolean query through a disjunction is likely to increase the number of blocks retrieved by the query. This said, please bear in mind that here the words added to the original query come from an external resource and may not be present in the collections queried.}. Table~\ref{tab:example_eldar} provides an example of these three types of boolean queries. Using the the \textit{eldar} package\footnote{\url{https://github.com/kerighan/eldar}}, we then computed, for very relevant and relevant documents, the number of blocks matching $q_{bool}$, those matching either $q_{bool}$ or $q^{syn}_{bool}$, and matching either $q_{bool}$ or $q^{rand}_{bool}$. The results are given in Table~\ref{tab:stats_matches} with the percentage increase with respects to the number of blocks matched by $q_{bool}$. One can observe that leveraging the knowledge about synonyms enables to match more blocks: up to 22.85\% increase in the number of blocks matched in the case of the Robust04 dataset (against 0.33\% for the random query), 14.70\% increase for GOV2 (against 0.26\% for the random query, 8.72\% in the case of MQ2007 (against 0.23\% for the random query) and 5.26\% increase for MQ2008 (against 0.16\% for the random query). 

\begin{table*}[t]
\centering
\caption{Statistics: number of blocks selected.}
\label{tab:stats_matches}
\resizebox{\linewidth}{!}{
\begin{tabular}{lrrrr}
\toprule
Dataset & MQ2007 & MQ2008 & Robust04 & GOV2 \\ \hline
\# of blocks matching $q_{bool}$ & 914,901 & 217,035 & 45,566 & 399,549 \\ 
\# of blocks matching $q_{bool}+q^{syn}_{bool}$ & 994,657 (+8.72\%) & 228,451 (+5.26\%) & 55,976 (+22.85\%) & 458,277 (+14.70\%)\\ 
\# of blocks matching $q_{bool}+q^{rand}_{bool}$ & 917,026 (+0.23\%) & 217,378 (+0.16\%) & 45,717 (+0.33\%) & 400,570 (+0.26\%)\\ 
\bottomrule
\end{tabular}
}
\end{table*}

\begin{figure*}[tb]
  \hspace{-0.8in}
  \begin{subfigure}[t]{0.3\textwidth}
      \centering
      \includegraphics[height=1.9in]{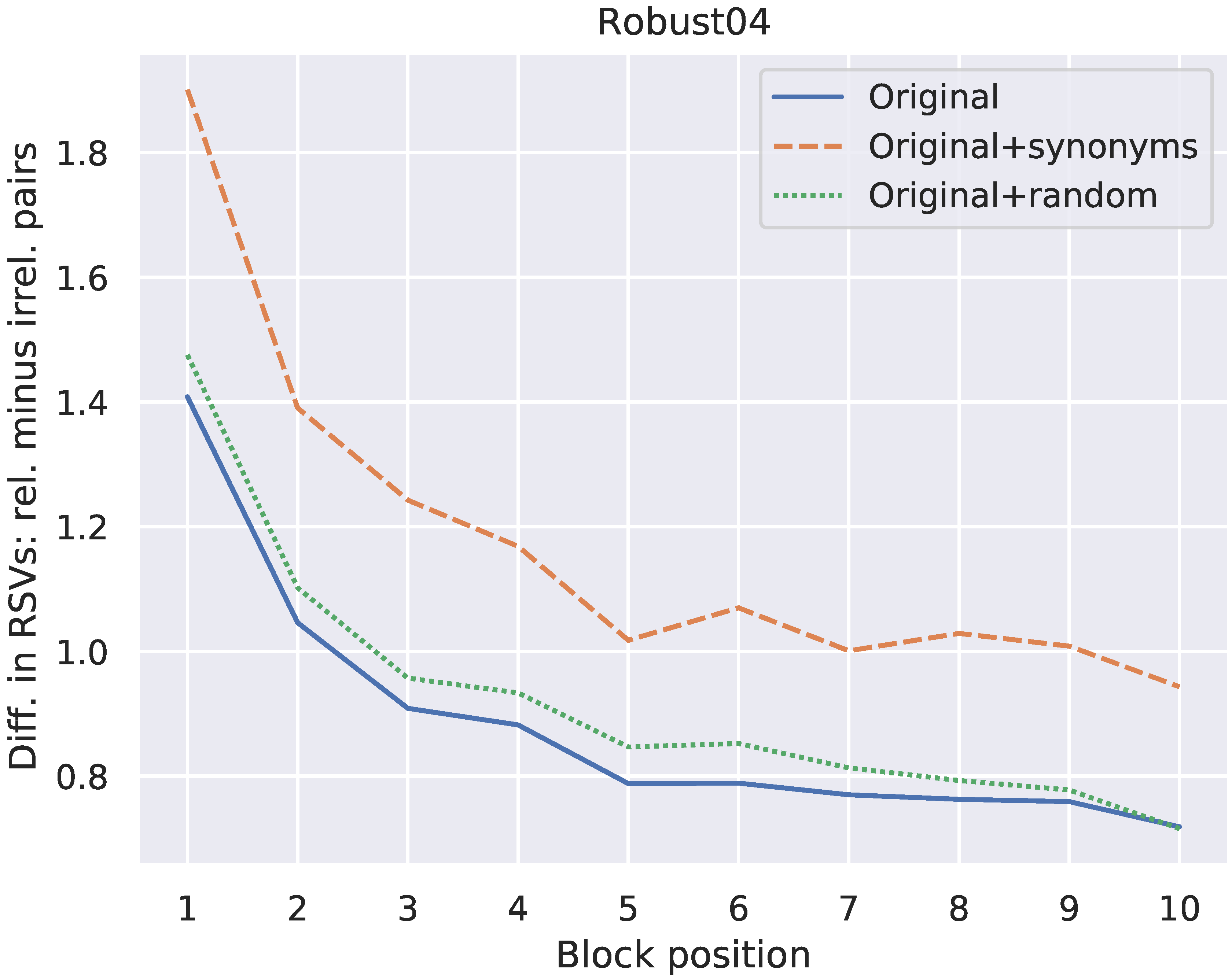}
      \caption{Difference of Robust04}
  \end{subfigure}%
  \hspace{1in}
  \begin{subfigure}[t]{0.3\textwidth}
    \centering
    \includegraphics[height=1.9in]{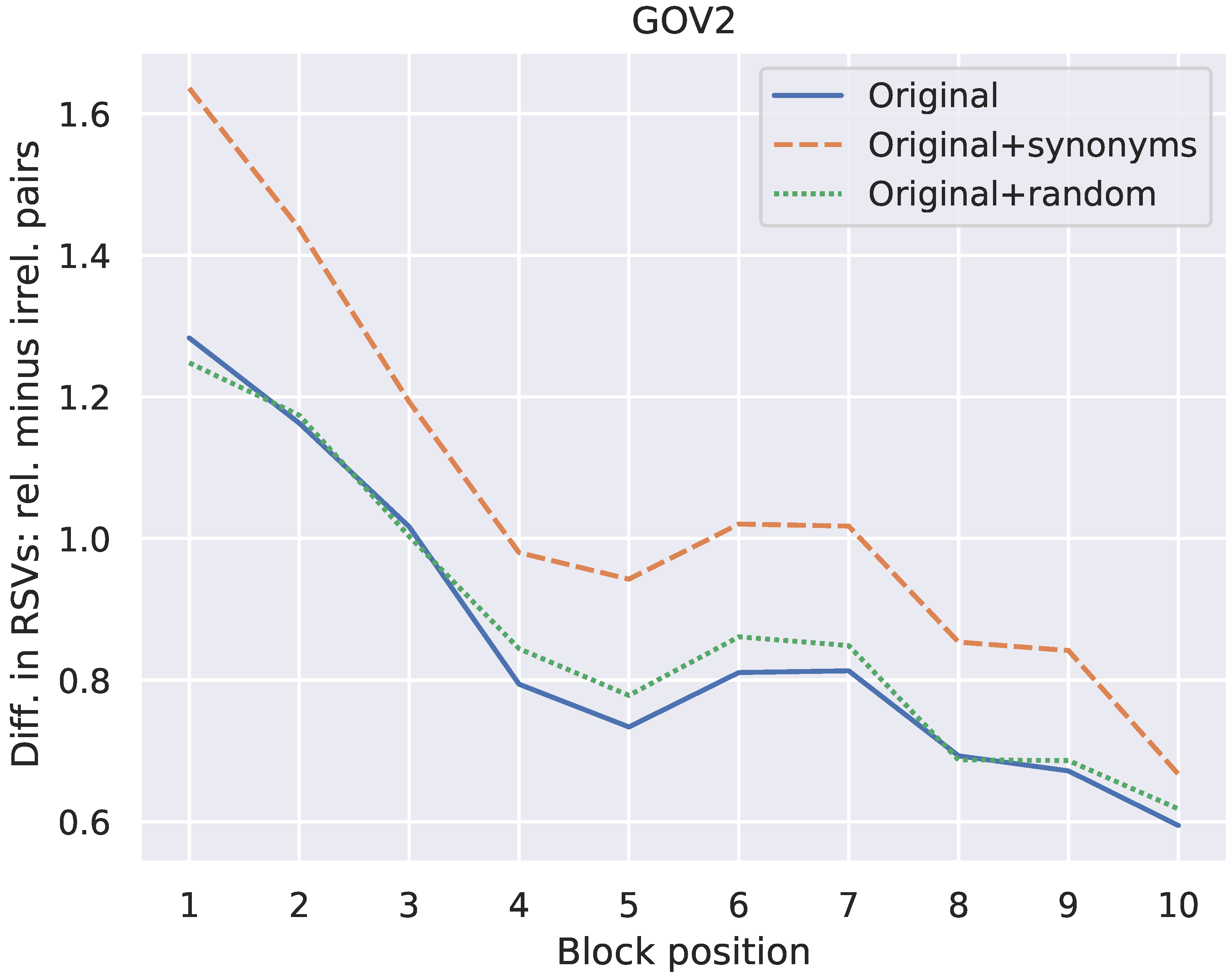}
    \caption{Difference of GOV2}
\end{subfigure}%

\hspace{-0.8in}
\begin{subfigure}[t]{0.3\textwidth}
  \centering
  \includegraphics[height=1.9in]{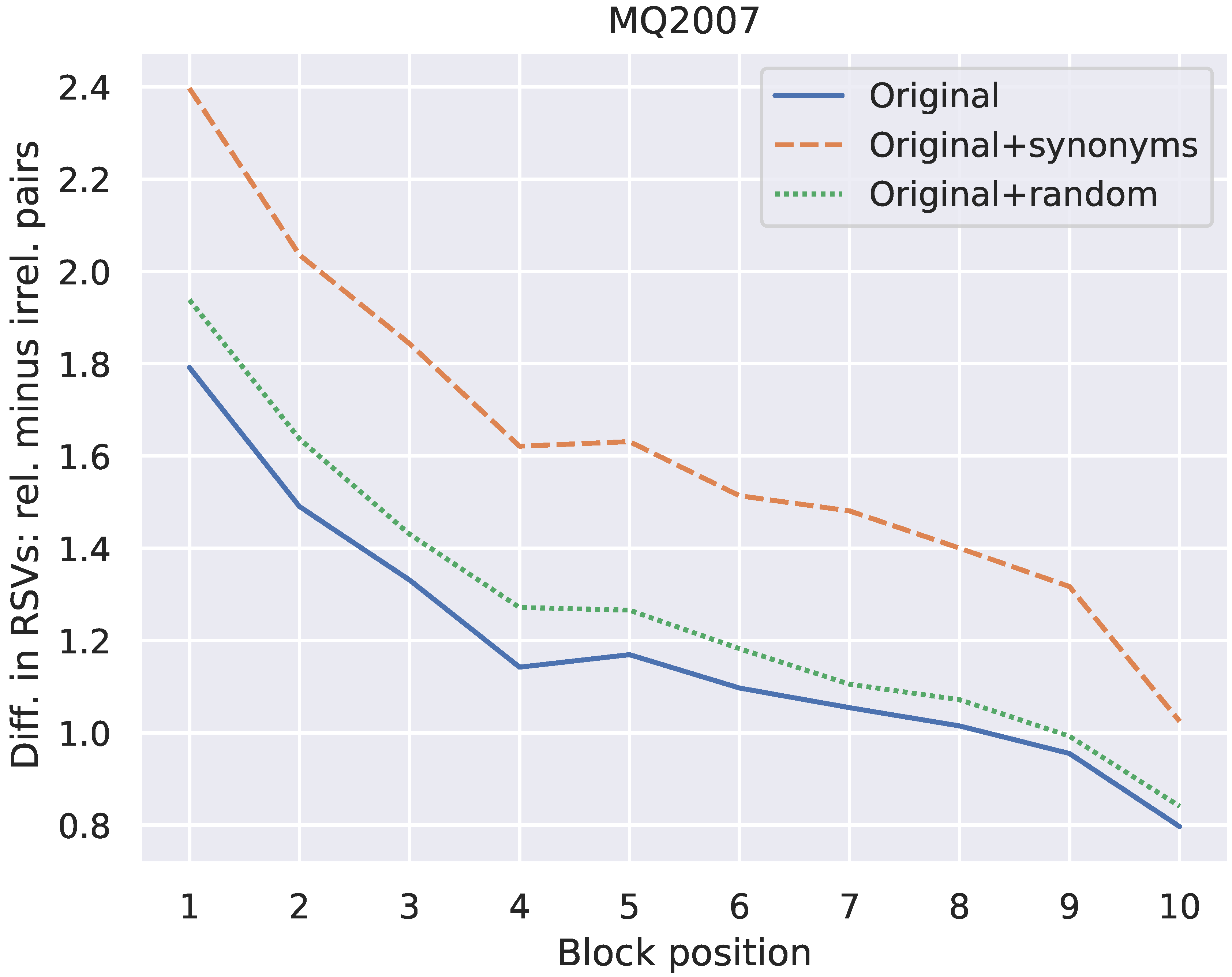}
  \caption{Difference of MQ2007}
\end{subfigure}%
\hspace{1in}
\begin{subfigure}[t]{0.3\textwidth}
  \centering
  \includegraphics[height=1.9in]{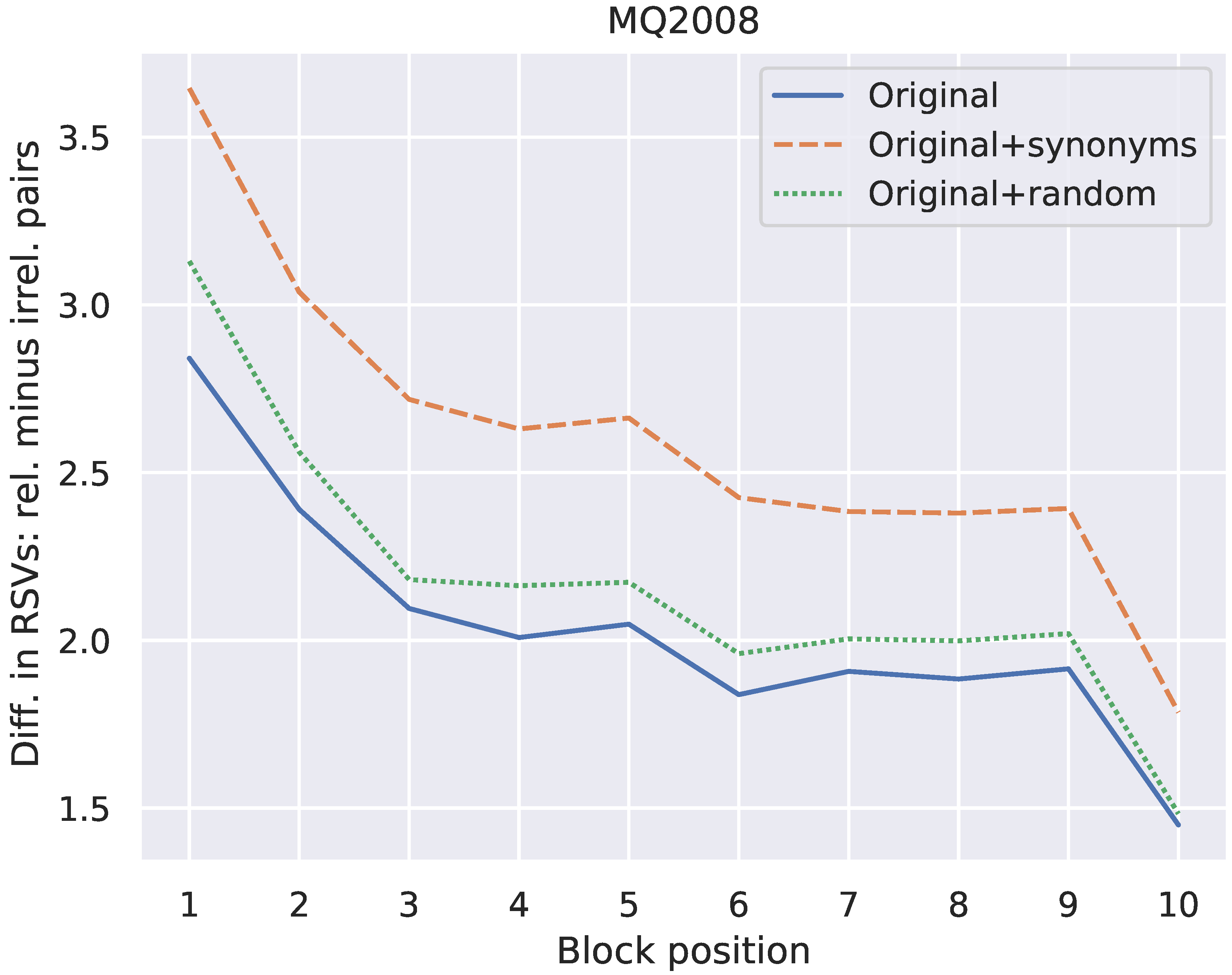}
  \caption{Difference of MQ2008}
\end{subfigure}%
  \caption{Difference in RSV scores between relevant and irrelevant documents for the original query $q$, the expanded one $q^{exp}$ and the random expanded one $q^{rand\_exp}$ across block positions.}
  \label{fig:diff_bm25_q_qqexp}
\end{figure*}

To complement the above analysis, we also assessed whether synonyms can leverage relevance signals in blocks at different positions. To do so, we used a standard expansion of the original query by simply adding all synonyms of all query terms without duplicates. For comparison purposes, we did the same with the words randomly selected in WordNet. The obtained queries, an example of which is given in Table~\ref{tab:example_eldar}, will be respectively referred to as $Original+synonyms$ and $Original + random$. We then computed the difference in average BM25 scores between relevant and irrelevant documents across all positions for the three types of queries: Original, Original + synonyms, Original + random. The results obtained are reported in Figure~\ref{fig:diff_bm25_q_qqexp}. As one can note, blocks in relevant documents score higher than the blocks in irrelevant documents, the gap being consistently and significantly higher when using the query with synonyms than when using the original query or the original query with additional random words. Furthermore, the two curves, Original and Original + random, are very close to each other and almost identical on GOV2. This shows that the increase in the BM25 scores when using synonyms is not due to the length of the query, and that synonyms help identify relevant signals in blocks. It thus may be useful to use matching strategies that go beyond an exact matching of query words.

In the next sections, we will present two different ways to select blocks in documents. The first one is based on standard IR functions, namely TF-IDF and BM25, to compute relevance scores between queries and blocks; it is thus based on an exact matching of words present in the query. The second one aims at learning a scoring function that exploits the semantic similarities between query words and document words. In both cases, the top scoring blocks are then used to compute the relevance score of the document.

\section{Selecting blocks with standard IR functions: TF-IDF and BM25}
\label{blockselectSec}

TF-IDF is a popular IR model which amounts to score a document through the product of the term frequency (TF) and inverse document frequency (IDF) scores of the query words present in that document. Applied at the block level, this yields the following retrieval status value (RSV): 
\begin{equation}
    RSV(q,b)_{TF-IDF} = \sum_{w \in q \cap b} \underbrace{(\ln{tf_w^b}+1)}_{TF} \cdot IDF(w). \nonumber
\end{equation}
in which $tf_w^b$ corresponds to the number of occurrences of word $w$ in block $b$ and $IDF(w)$ to the inverse document frequency  of word $w$. 
$IDF(w)$ is defined here according to scikit-learn \cite{pedregosa2011scikit} by:
\begin{equation}
\label{idf}
    IDF(w)=\ln \frac{N+1}{df_{w}+1}, \nonumber
\end{equation}
where $N$ is the number of documents in the collection and $df_{w}$ corresponds to the number of documents containing $w$.

For BM25 \cite{RobertsonZ09,robertson2009probabilistic}, the RSV score is defined by:
\begin{equation}
    RSV(q,b)_{BM25} = \sum_{w \in q \cap b} IDF(w) \cdot \frac{tf_{w}^b}{k_1 \cdot (1-b+b\cdot \frac{l_{b}}{l_{avg}})+tf_{w}^b}, \nonumber
\end{equation}
where $l_{b}$ is the length of block $b$, $l_{avg}$ the average length of the blocks in $d$, and $k_1$ and $b$ two hyperparameters. As standard in this setting, we use the IDF formulation of Lucene version as shown in \citet{kamphuis2020bm25} which slightly differs from the previous one:
\begin{equation}
    IDF(w)=\ln \frac{N+1}{df_{w}+0.5} . \nonumber
\end{equation}

According to \cite{nottelmann2003retrieval}, RSVs in different IR systems have different scales. In the boolean model, RSVs are either 0 or 1. Vector space model can generate RSV in $[-1,1]$ by cosine similarity or scalar product $\mathbb{R}$. In this section, we rely on TF-IDF and BM25, which can be viewed as an model that rely on term matching and the scale of the RSVs is $\mathbb{R}$. The blocks are ranked according to RSVs by descending order.

\paragraph{Block or document level IDF} As the reader may have noticed, the IDF is based on documents and not on blocks. There are two main reasons for this. First of all, considering blocks instead of documents for the IDF may artificially increase the $df$ score of a word since important words of a document are likely to occur in many blocks of that document. The second reason is that one can directly re-use existing IDF scores computed at the document level. Note that all words are lowercased prior to compute TF-IDF and BM25 scores.

\begin{figure}[!t] \centering \includegraphics[]{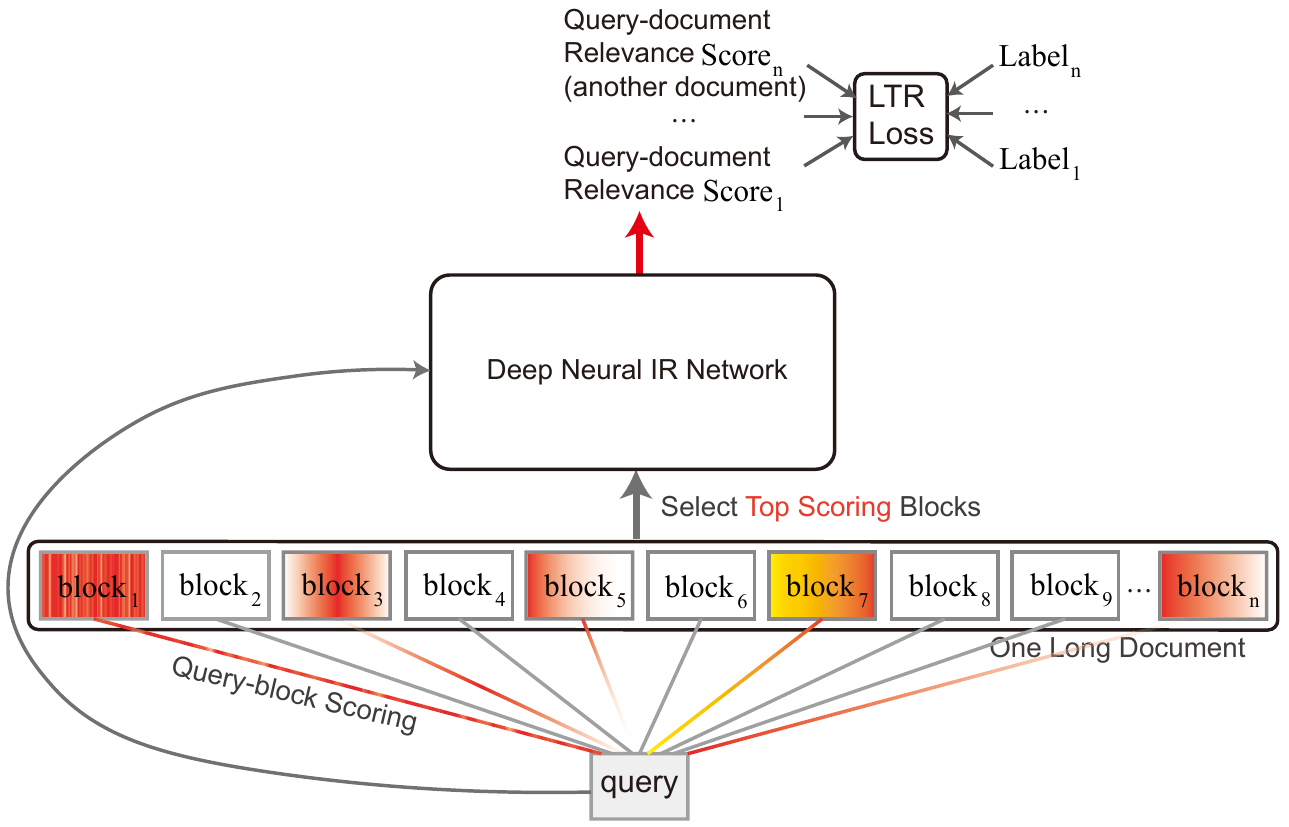}
\caption{An illustration of the architecture of KeyB (e.g., TF-IDF or BM25).}
\label{archi} 
\end{figure}

\begin{figure*}[tb]
    \centering
    \begin{subfigure}[t]{0.3\textwidth}
        \centering
        \includegraphics[height=1.5in]{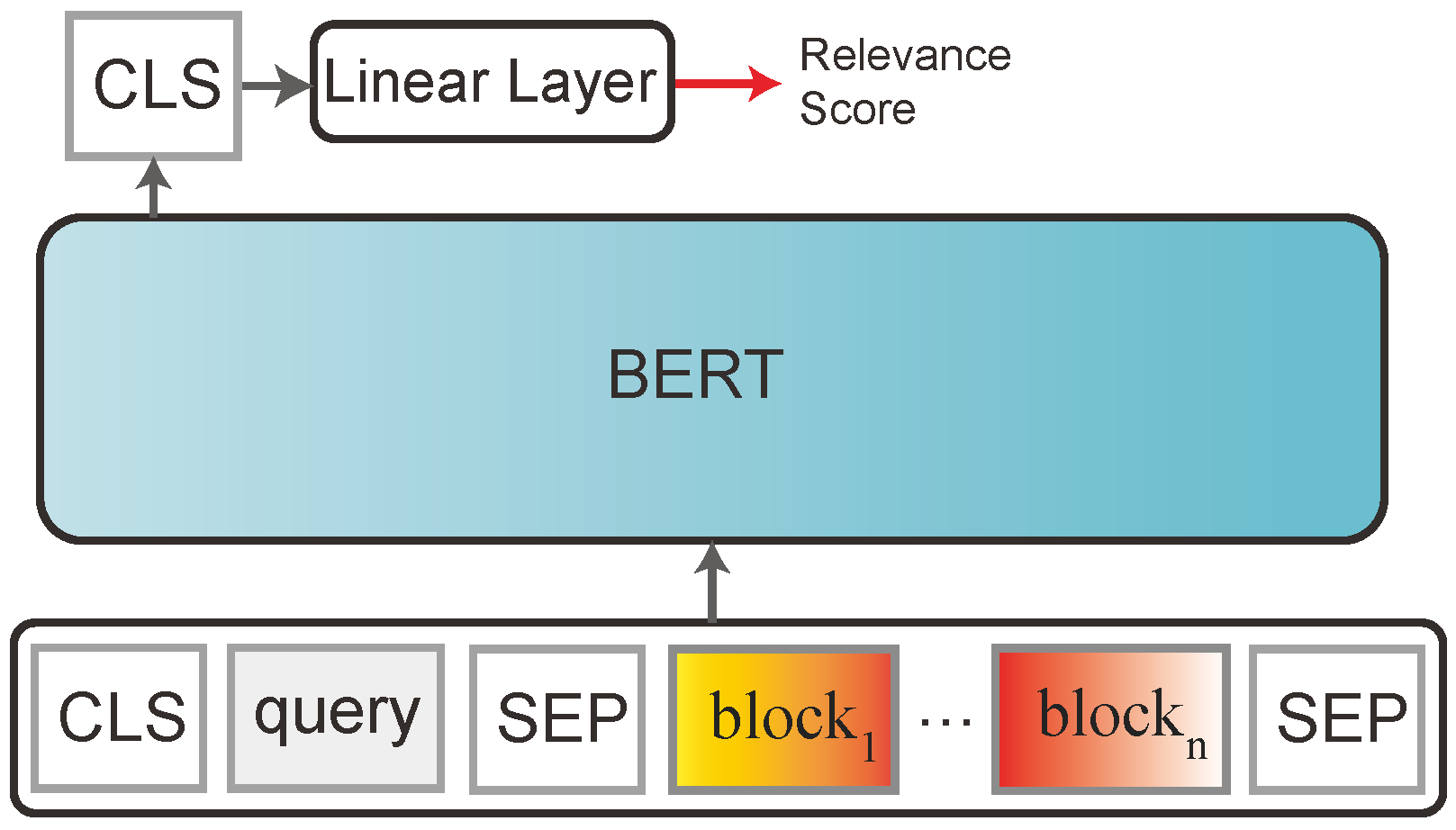}
        \caption{Vanilla BERT neural IR network}
        \label{fig.vbert}
    \end{subfigure}%
    \hspace{1.5in}
    \begin{subfigure}[t]{0.3\textwidth}
        \centering
        \includegraphics[height=1.6in]{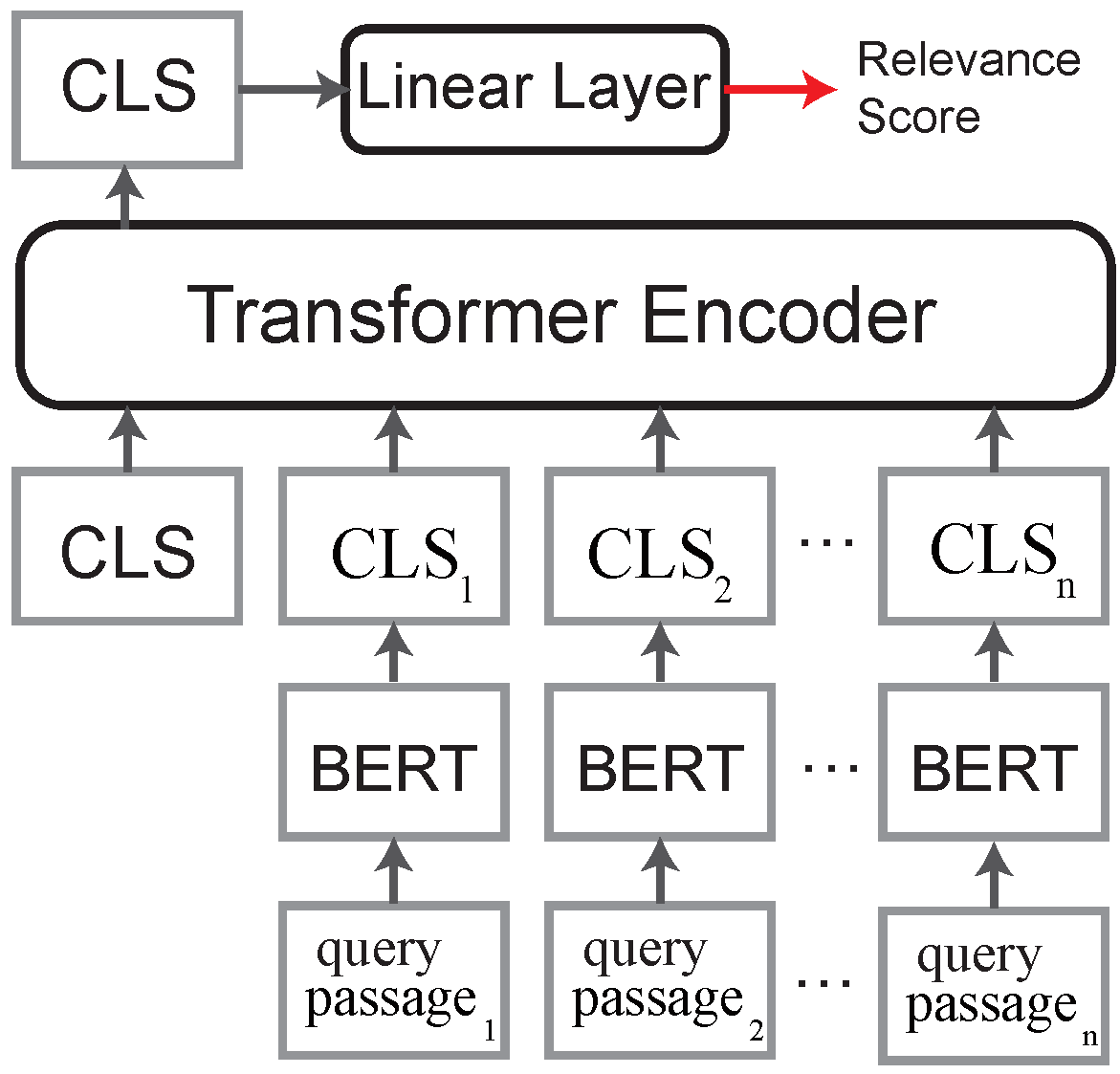}
        \caption{PARADE neural IR network}
        \label{fig.parade}
    \end{subfigure} %
    \caption{Different deep neural IR networks.}
    \label{fig.dnn}
\end{figure*}

The overall architecture of an IR neural system relying on the above standard IR models to select blocks is presented in Figure~\ref{archi}. As one can note, the query-block scoring part is used to select relevant blocks across the whole document, which can be viewed as a pre-ranking strategy. The Deep Neural IR Network can represent any neural IR network which generates relevance scores for query-document pairs, scores which can in turn be used as input to a learning-to-rank (LTR) loss, be it a pairwise or listwise loss. We focus in this study on two state-of-the-art neural IR Models, namely Vanilla BERT, in which case we refer to the models obtained as \textit{KeyB(vBERT)}$_{TF-IDF}$ and \textit{KeyB(vBERT)}$_{BM25}$, and PARADE, in which case we refer to the models obtained as \textit{KeyB(PARADE$k$)}$_{TF-IDF}$ and \textit{KeyB(PARADE$k$)}$_{BM25}$, $k$ representing in that case the number of passages retained. Figure~\ref{fig.dnn} provides an illustration of Vanilla BERT and PARADE.

\subsection{KeyB(vBERT)}
\label{keybvbertsec}

The KeyB(vBERT) models rely on four steps:
\begin{enumerate}
    \item \textit{Block segmentation} We adopt here the dynamic programming method proposed in \cite{NEURIPS2020_96671501} to segment documents into blocks, each block having a maximum of 63 tokens. This method sets different costs for different punctuation marks and aims at segmenting in priority on strong punctuation marks such as "." and "!". It is thus close to a sentence segmentation procedure.
    \item \textit{Block selection} Each block is assigned a relevance score provided by either TF-IDF or BM25, as described above.
    \item \textit{Query-blocks representation} The most relevant blocks are then concatenated together in their order of appearance in the document and with the query to form the input of BERT (see Figure~\ref{fig.dnn}). As the input number of tokens for BERT is limited to 512, the last block is truncated if necessary. The number $n$ of selected blocks depends on the capacity of BERT and is defined by:
\begin{equation}
    3+l_q+\sum_{i=1}^{n-1} l_{b_{i}}+\mbox{trunc}({l_{b_{n}}})=512, \nonumber
\end{equation}
where $l_q$ denotes the length of the query and $l_{b_{i}}$ the length of a block, potentially truncated for the last selected block. The final query-representation corresponds to the [CLS] embedding of the learned BERT.
    \item \textbf{Document ranking} We rely here on a one-layer dense, linear network to generate the final relevance score, using a learning to rank loss computed on a mini-batch (see Section~\ref{lossfuncsec}).
\end{enumerate}

\subsection{KeyB(PARADE$k$)}
\label{improveParadeSec}

As mentioned in Section \ref{sec:related-work}, PARADE is a state-of-the-art model which computes a query-document representation on the basis of the query-passage representations. A document is first segmented into passages. Each passage is then fed, together with the query, to a BERT model. The CLS embedding thus obtained corresponds to the query-passage representation. Denoting by $p_i$ the $i^{th}$ passage and $p_i^{cls}$ the corresponding query-passage representation, one has:
\begin{equation}
    p_i^{cls}= BERT(q, p_i) . \nonumber
\end{equation}
The query-passage representations are then aggregated to obtain the query-document representation, that is finally fed to a feed-forward neural network to obtain the relevance score of the document. 

Four types of aggregation methods have been proposed: PARADE–Max, PARADE–Attn, PARADE–CNN and PARADE–Transformer. PARADE–Max uses a max pooling operation on the passage relevance representations. PARADE–Attn assumes that each passage contributes differently to the relevance of a document to the query and passage weights are predicted by a feed-forward network. PARADE–CNN stacks Convolutional Neural Network (CNN) layers in a hierarchical way whereas PARADE–Transformer stacks a full attention model which enables query-passage representations to interact in a hierarchical manner through a transformer. Its architecture is depicted in Fig.~\ref{fig.parade}. Because of its good behavior \cite{li2020parade}, we have retained this last aggregation method here. We will refer to is as just PARADE in the remainder.

Let $x^{(\ell)}$ denote the input of the $\ell$ transformer layer. $x^{(0)}$ consists in the concatenation of the query-passage representations ($ p_i^{cls}$). $x^{(\ell+1)}$ is then given by:
\begin{align}\label{eq.PARADE_transformer}
    x^{(\ell+1)} = \operatorname{LayerNorm} (h + \operatorname{FFN}(h)), \nonumber\\
    \text{with} \,\, h = \operatorname{LayerNorm} (x^{(\ell)} + \operatorname{MultiHead} (x^{(\ell)})). \nonumber
\end{align}
LayerNorm refers to the layer-wise normalization described in \cite{ba2016layer} and MultiHead to the multi-head self-attention \cite{vaswani2017attention}. FFN is a two-layer feed-forward network with a ReLU activation in between. The [CLS] vector of the final output layer, which is denoted by $d^{cls}$, constitutes the query-document representation. A linear layer is then used to generate the query-document relevance score:
\begin{equation}
    RSV(q,D) = Wd^{cls},\nonumber
\end{equation}
where $W$ is a learnable weight matrix.

It is important to note that the PARADE model described above can have a high complexity when the number of passages considered is large (for example, when using 10 passages, the model can not fit on a standard GPU with 11 GB memory even with mixed precision). Indeed, in that case, the input consists in a large tensor which can only be stored in a large CUDA memory \cite{gao2020estimating} with high computational complexity. To avoid this, the number of passages is limited to $16$. When a document contains more than 16 passages, then only the first passage, the last passage and $14$ sampled passages are used. Whether one restricts the number of passages or not, one problem faced by PARADE lies in the fact that non relevant passages can bring noise in the query-document representation, and hamper the computation of a precise retrieval score. To address this problem, we propose here to select a fixed, small number of passages, denoted by $k$. The selected passages are then concatenated and fed to a transformer as in the original PARADE model. When using standard IR functions as described above to select passages, the query-document representations denoted by $d_{TF-IDF}^{cls}$ and $d_{BM25}^{cls}$ are thus obtained. The final relevance score of the document is then obtained by: 
\begin{equation}
    RSV(q,D) = Wd_{TF-IDF}^{cls} \text{ or } Wd_{BM25}^{cls}.\nonumber
\end{equation}

\subsection{Model training}
\label{lossfuncsec}

The block/passage selection process is applied in both the training and testing phases. The BERT models used in Vanilla BERT and PARADE are fine-tuned during training. The parameters of all the other components (final layer for Vanilla BERT, final layer and transformer layers for PARADE) are directly trained. For fine-tuning and training, the following pairwise hinge loss \cite{guo2016deep} is used:
\begin{equation}
\mathcal{L}(q,d^+,d^-;\Theta) = \max(0, 1-s(q, d^+;\Theta)+s(q, d^-;\Theta)), \nonumber
\end{equation}
where $q$ represents a query, $(d^+_q,d^-_q)$ a positive and negative training document pair for $q$, $\Theta$ the parameters of the model considered and $s$ the predicted relevance score for a query-document pair. Other choices are of course possible, including ones based on a listwise loss. We chose the pairwise hinge loss here as it is a very common choice, used in many IR methods \cite{macavaney2019cedr, li2021keybld,pang2017deeprank}.

\section{Learning to select blocks}
\label{keyb3Sec}

The analysis conducted in Section~\ref{sec:analysis} suggests that a fuzzy matching procedure may be preferred over one based on an exact matching. We thus aim here to learn a scoring function that exploits the semantic similarities between words in queries and blocks using the same two neural IR model as before, Vanilla BERT and PARADE.

\subsection{Improving Vanilla BERT}

\begin{figure}[!t] \centering \includegraphics[width=0.95\textwidth]{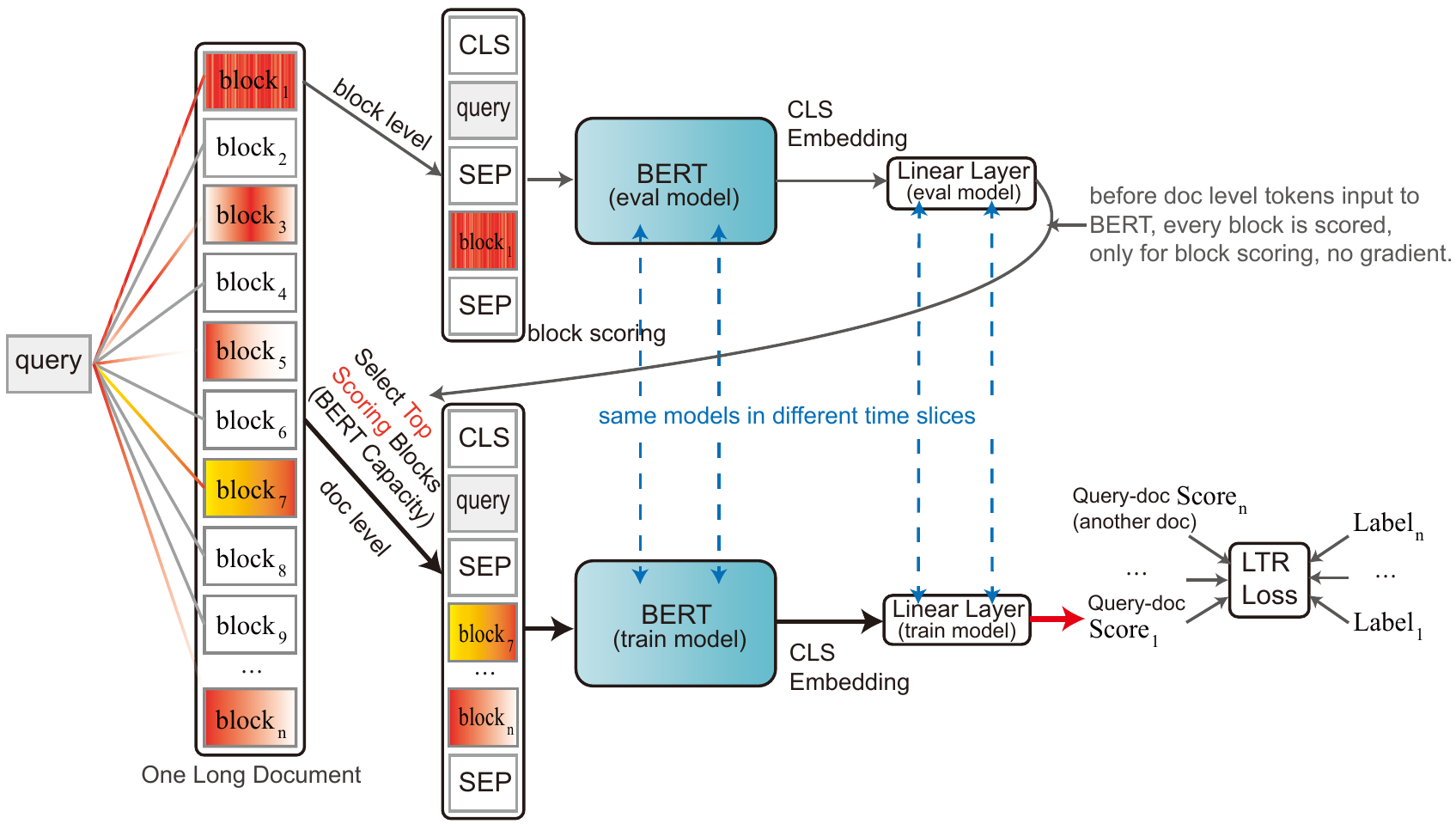}
\caption{An illustration of the architecture of KeyB(vBERT)$_{BinB}$. Here, the BERT model and linear layer are used to select blocks too. While the neural model being trained with document level annotations, this model would become able to score blocks for a query.}
\label{archib3} 
\end{figure}

We focus here on the Vanilla BERT model to compute the relevance score of a block. That is, instead of just using the semantic representation of query-block pairs to compute the relevance score of a document as in the previous models, we also make use here of these representations to select the most appropriate blocks. Furthermore, we propose here to share the semantic representation for both purposes, selecting blocks and computing the overall relevance score, as both are based on the relevance information contained in each block. It is of course possible to make use of two different models, with however higher computational and training costs. 
The overall architecture of the model proposed is depicted in Figure~\ref{archib3}, in which the same BERT model and linear layer are used at different time slices, first to compute a query-block representation, from which ([CLS] embedding) the relevance score of the block is derived, and then to compute the query-document representation ([CLS] embedding) based on the top ranked blocks, finally to obtain the score of the document. This query-document representation is identical to the one used in the KeyB(vBERT) model, the only difference lying in the way the blocks are selected. For the query-block represntation, both the BERT model and the linear layer are only used for scoring and are not updated via back-propagation (hence the term "eval model" used in Figure~\ref{archib3}). The score of a block $b$ for a given query $q$ is defined by:
\begin{equation}
    RSV(q,b)_{BERT} = {W_d}b^{cls},\nonumber
\end{equation}
where $W_d$ is the weight of the dense linear layer on top of BERT, and $b^{cls}$ is the query-block relevance representation:
\begin{equation}
    b^{cls} = BERT(q,b).\nonumber
\end{equation}

For a given query, the BERT model is first used to generate a relevance score for each block in the document. Since the BERT model is not well fine-tuned at the beginning of the process, the block selection process acts as an almost random selector. However, with the query-document level relevance labels, after each back-propagation, the BERT model is improved and provides better relevance scores for each block in the next iterations. In return, the BERT model benefits from the block selection too. This can be viewed as a self-evolution process: the BERT model evolves to provide appropriate query-block representations to be able to select blocks, and meanwhile, appropriate query-document representations to be able to generate relevance scores for query-document pairs. Because of this self-evolution, we refer to this model as KeyB(vBERT)$_{BinB}$, where ${BinB}$ means ''BERT in BERT''.

\subsection{Improving PARADE}
\label{keybParadeBibSec}

\begin{figure}[!t] \centering \includegraphics[width=0.8\textwidth]{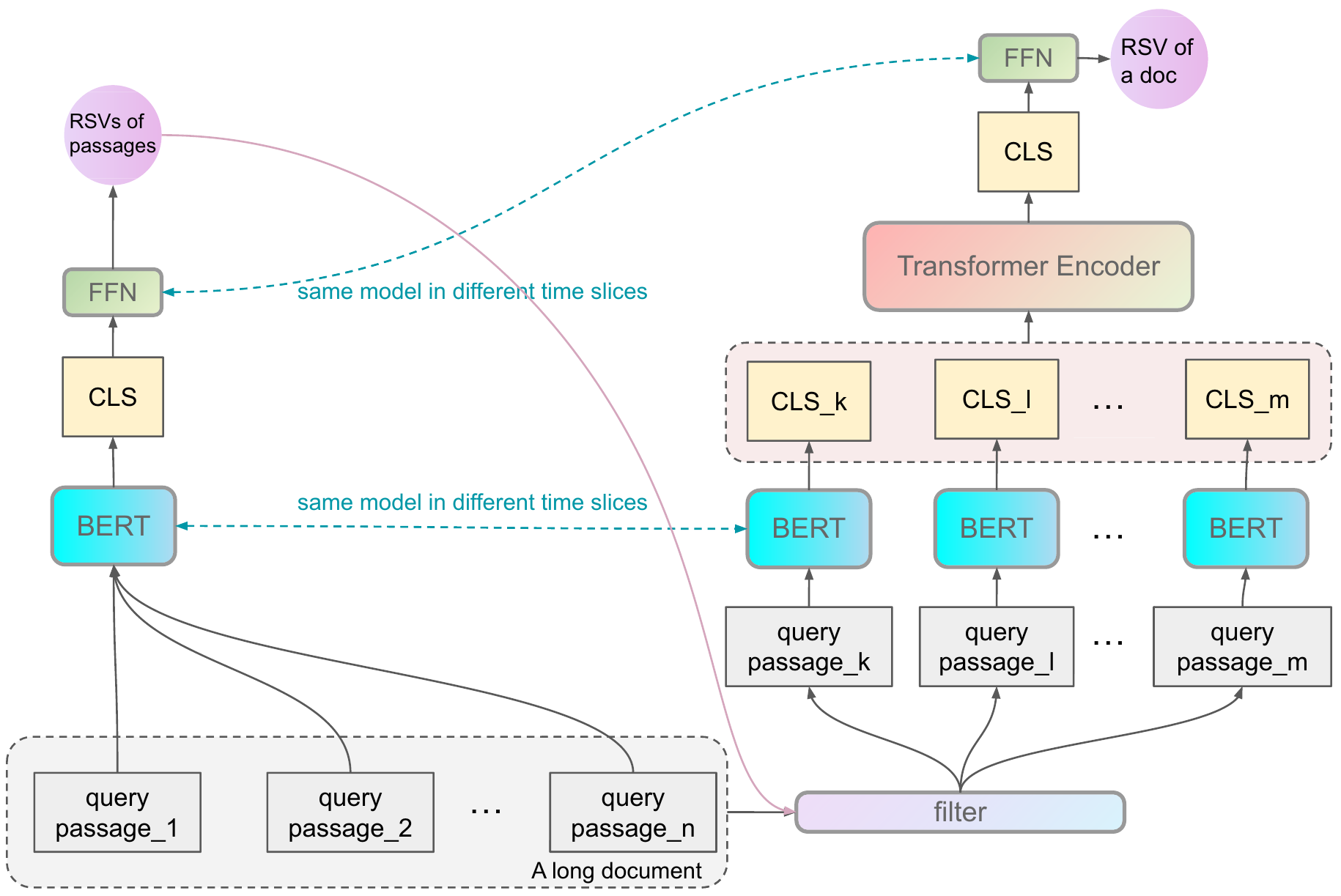}
  \caption{An illustration of the architecture of KeyB(PARADE$k$)$_{BinB}$. Here, the BERT model and linear layer of PARADE are used to select blocks too.}
  \label{archikeybParadeBib} 
  \end{figure}
  
We here propose to improve the PARADE model with the learning approach for selecting passages.

To do so, we first follow the same strategy as the one used for Vanilla BERT by trying to use the same modules for both selecting passages and scoring documents. A major difficulty for doing so here is that the Transformer encoder used in PARADE to score a complete document takes as input the representation of several query-block pairs. As such, it cannot be used to score a single block. For this reason, we first propose to only share the initial BERT model and the final feed-forward neural network as illustrated in Figure~\ref{archikeybParadeBib}. In this new architecture, one obtains the [CLS] query-passage representation, denoted by $p^{cls}$ for passage $p$, using the base BERT model of PARADE which is shared in different time slices. The final feed-forward neural network is used to provide, from a CLS representation, a score for a query-passage pair (Fig.~\ref{archikeybParadeBib}, left) and a score for a query-document pair, where a document is seen as the concatenation of passages (Fig.~\ref{archikeybParadeBib}, right). The score of a passage $p$ for a given query $q$ is then defined by:
\begin{equation}
  RSV(q,p)_{BERT} = {W_d}p^{cls},\nonumber
\end{equation}
where $W_d$ is the weight of the feed-forward neural network after the Transformer encoder, and $p^{cls}$ is the query-passage relevance representation:
\begin{equation}
  p^{cls} = BERT(q,p).\nonumber
\end{equation}
We refer to this approach as KeyB(PARADE$k$)$_{BinB}$, where $k$ is the number of selected passages.

\begin{figure}[!t] \centering \includegraphics[width=0.8\textwidth]{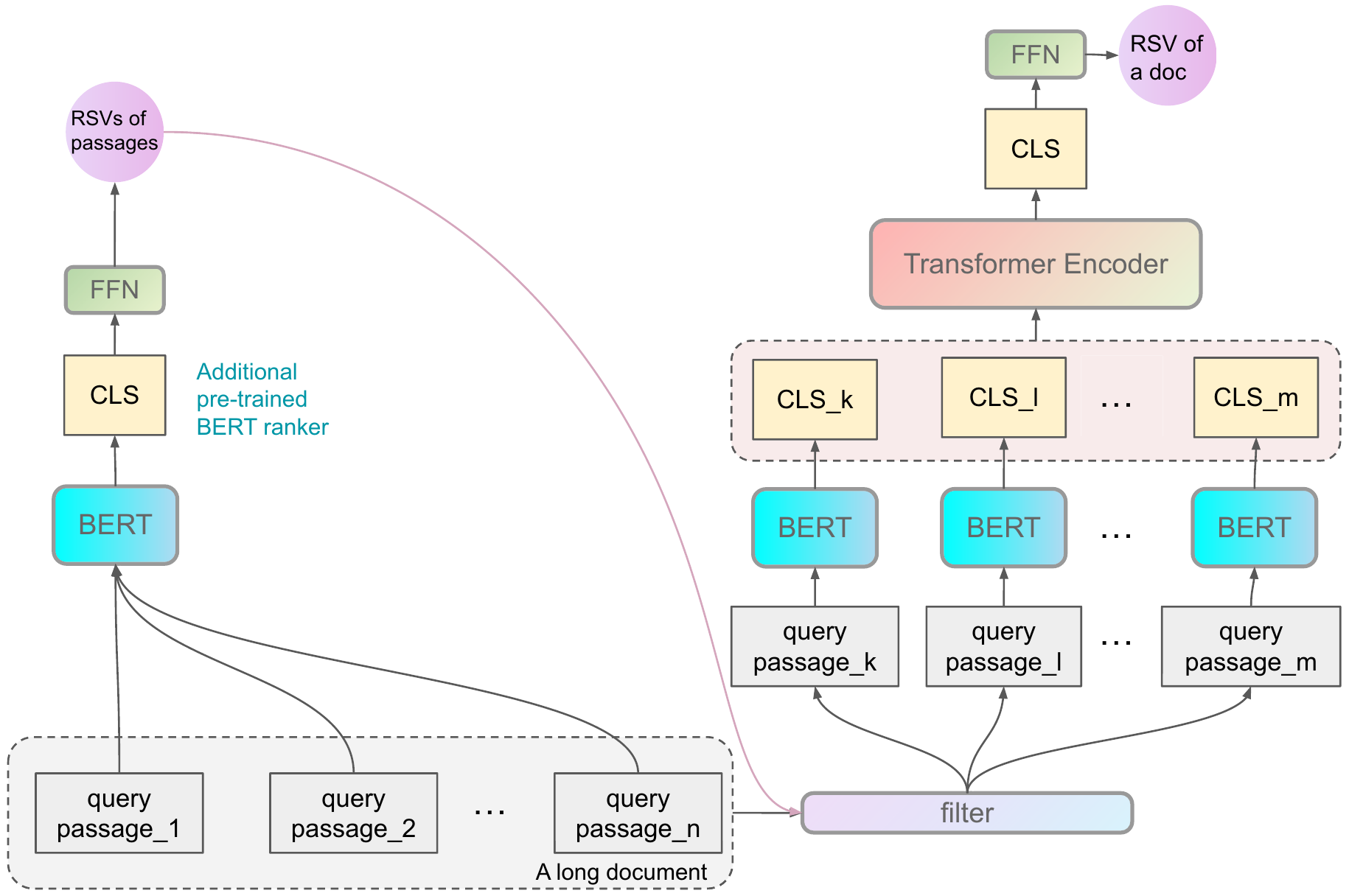}
  \caption{An illustration of the architecture of KeyB(PARADE$k$)$_{BinB2}$. Here, additional BERT model and additional linear layer are used to select blocks.}
  \label{archikeybParadeBib2} 
\end{figure}

If the previous attempt makes it possible to reuse modules for selecting passages and scoring documents, it may however suffer from the fact that the same feed-forward neural network is required to produce a relevance score from two different CLS representations: one restricted to a single query-passage pair for selecting passages, and one resulting from an encoding, through a Transformer, of several query-passage representations for scoring the document. 
We propose to fix this issue by decoupling the passage scoring module from the main model, as illustrated in Figure~\ref{archikeybParadeBib2} which relies on a different BERT module and feed-forward neural network to select passages. But which BERT model and feed-forward layer to use? A simple answer to this question is to re-use the pretrained BERT ranker and final feed-forward neural network of one of the KeyB(vBERT) models previously proposed as these models are compatible with the input used here. The score of a passage $p$ for a given query $q$ is in this case defined by:
\begin{equation}
  RSV(q,p)_{BERT2} = {W_{d}^2}p^{cls_2},\nonumber
\end{equation}
where $W_{d}^2$ is the weight of the feed-forward neural network following the additional BERT module, and $p^{cls_2}$ is the query-passage relevance representation after the additional BERT module:
\begin{equation}
  p^{cls_2} = BERT2(q,p).\nonumber
\end{equation}
In practice, we use the BERT module and feed-forward neural network from the KeyB(vBERT)$_{BM25}$ model and refer to this approach as KeyB(PARADE$k$)$_{BinB2}$, where $k$ is the number of selected passages.

\section{Experiments on standard IR collections}
\label{sec:experiments}


We conducted a first series of experiments on the same collections as the ones used in the analysis presented in Section~\ref{sec:analysis}, namely MQ2007, MQ2008, Robust04 and GOV2. These experiments aim at assessing the validity of the models proposed, KeyB(vBERT) and KeyB(PARADE$k$)\footnote{These models are developed on top of the Georgetown IR Lab implementation and are available at: \url{https://github.com/lmh0921/keyB}.}, and at comparing them to models that have provided state-of-the-art results on these collections. These latter models are:
\begin{itemize}
\item \textbf{DeepRank}: We compare with the DeepRank with CNN based measure network, which leads to better results than 2D-GRU based on \cite{pang2017deeprank}.  We used the PyTorch implementation of DeepRank \footnote{\url{https://github.com/pl8787/DeepRank_PyTorch}} with given hyperparameter for the architecture. Following the implementation, the number of words per document is set to 2000. Hence, by adopting the default parameters the documents longer than 2000 tokens are truncated.
\item \textbf{PARADE}: We compare our model with the PARADE-Transfomer version as this version performs mostly better and we call this baseline PARADE in short. This method is a state-of-the-art IR method which first segments documents into passages that are fed to a BERT model. Transformer layers are then used to compute global attention scores over the [CLS] embeddings of different passages. A final linear layer is used for computing the document level relevance score. We have used the open-sourced PyTorch implementation\footnote{\url{https://github.com/capreolus-ir/capreolus}}.
\item \textbf{CEDR-KNRM}: CEDR is also a reported state-of-the-art model that incorporates BERT’s classification vector into existing neural models. We choose the CEDR-KNRM as the baseline for it is reported better than other variants. We have used the Georgetown IR Lab implementation\footnote{\url{https://github.com/Georgetown-IR-Lab/cedr}} with the Hugging Face transformer module\footnote{\url{https://github.com/huggingface/transformers}}.
\end{itemize}

For assessing the various components of the models proposed, we also used the following baseline models:
\begin{itemize}
\item \textbf{BM25}: We use the BM25 implementation of Anserini \cite{yang2018anserini}, with default hyperparameters. This model is the one presented in Section~\ref{blockselectSec} and is used both as a baseline on all collections. In addition, it is used as a first stage ranker, retaining only the top 200 documents, for the neural IR models on Robust04 and GOV2.
For MQ2007 and MQ2008, we converted the original documents to JSON files which can be indexed by Anserini. We then used the BM25 model from Anserini as a baseline but not as a first stage ranker as in MQ2007 and MQ2008 each query only contains dozens of labeled documents for ranking. 
For all experiments, the hyper-parameters $k1$ and $b$ of BM25 are set to Anserini's default values: 0.9 and 0.4.
\item \textbf{Vanilla BERT}: This is a BERT baseline that truncates long documents to the first 512 tokens. Except for this difference in the input used for BERT, the architecture is the same as the one of KeyB(vBERT). This baseline thus allows one to evaluate the impact of key blocks. For implementation, we have used the same library as the one for CEDR-KNRM.
\item \textbf{Random Select}: This architecture is the same as the one with KeyB(vBERT) except that it does not incorporate the block selection mechanism. To be specific, a long document is also firstly segmented into blocks, but each block is given a random score. This is to say, without local block pre-ranking step, the blocks are selected randomly. 
\end{itemize}

\subsection{Experimental design}
\label{experDesignSec}

It is common in IR to first filter documents with a classical IR system prior to re-rank them with a more complex (and usually more time consuming) system \cite{nogueira2019passage,macavaney2019cedr,dai2019deeper,li2020parade}. We adopt here this approach and use BM25 as the first filtering system, retaining only, for each query, the first 200 documents on Robust04 and GOV2 as done in \cite{macavaney2019cedr}. As queries in MQ2007 and MQ2008 are associated with far less than 200 documents, this filtering step is not necessary. Following \cite{pang2017deeprank}, we merged the MQ2007 and MQ2008 training sets as the training set of MQ2008 is relatively small. The validation and testing sets remain unchanged. For all neural IR models, the pairwise hinge loss \cite{pang2017deeprank} is used for training the models.

Furthermore, for all experiments, 5-fold cross-validation is used with three folds for training, one fold for validation and one fold for testing. For Robust04 and GOV2, we used the keyword (title) version of queries \cite{dai2019deeper}. For MQ2007 and MQ2008, the default queries are used. All neural IR models based on BERT use the “BERT-Base, Uncased, L=12, H=768”\footnote{\url{https://github.com/google-research/bert}} pre-trained language model (but not further pre-trained with additional data) for fair comparison. 

For DeepRank, we followed the experimental setup of \cite{pang2017deeprank} and used GloVe embeddings \cite{pennington2014glove} of dimension $50$, which are pre-trained on Wikipedia 2014 + Gigaword 5. For the preprocessing of document and query words, we applied lower-case, removed English stop words, stemmed with Krovetz stemmer \cite{krovetz2000viewing} and removed words occurring less than 5 times. The Adam optimizer is used for training the network and the learning rate is searched over values of $\{0.01, 0.005, 0.001\}$. We selected the model that leads to the best MAP score on the validation set.

For PARADE, the first and last passages are always selected and the other passages are randomly sampled\footnote{As done \url{https://github.com/canjiali/PARADE/blob/master/generate_data.py} in line 304.}.
 The number of passages considered is a hyperparameter in PARADE that needs to be set. Following \cite{li2020parade}, the passages are obtained with 225 document tokens with stride size 200, the maximum passage sequence length is set as 256, and the number of passages is set to 16 for all collections. Note again that for a fair comparison, BERT is not further trained on MS-MARCO \cite{nguyen2016ms}. For all BERT based IR models, we use the BERT implementation of PyTorch Huggingface library \cite{wolf2019huggingface}.

For the variants of PARADE we have proposed in Section \ref{improveParadeSec}, we have used TF-IDF or BM25 to select the top 5 passages. The choice of 5 passages provides a good balance for effectiveness, as ca. 1000 tokens are considered, and efficiency, as a standard RTX 2080ti GPU with 11GB memory is not able to deal with 12 passages and automatic mixed precision \cite{micikevicius2018mixed} for example. The resulting models, based on PARADE and integrating the top 5 passages using TF-IDF or BM25 for local pre-ranking, are referred to as KeyB(PARADE5)$_{TF-IDF}$ and KeyB(PARADE5)$_{BM25}$. The variants of PARADE proposed in Section \ref{keybParadeBibSec} and with top 5 passages are are referred to as KeyB(PARADE5)$_{BinB}$ and KeyB(PARADE5)$_{BinB2}$.
We here use the "cross-encoder/ms-marco-MiniLM-L-12-v2"\footnote{\url{https://www.sbert.net/docs/pretrained_cross-encoders.html}} version model as the standalone BERT ranker for KeyB(PARADE5)$_{BinB2}$. This standalone BERT ranker is firstly finetuned on each collection as the way of KeyB(PARADE5)$_{BM25}$ (trained with document level labels), then it can generate a RSV for each query-passage pair (this means that a query and passage are concatenated and directly input to the BERT ranker).
 All other settings are the same as the ones for the PARADE model described above. For comparison purposes, we also used another variant of PARADE, called PARADE5, relying on the first, last and 3 randomly chosen passages.

Each model is trained for a maximum of 10 epochs. For Robust04 and GOV2, one epoch represents 1024 batches of 
two pairs, each pair being of the form ${((q,d_q^+),(q,d_q^-))}$ 
where $d_q^+$ is a positive document for query $q$ and $d_q^-$ a negative document. Since MQ2007 and MQ2008 have more queries than Robust04 and GOV2, each epoch of these two collections is composed of 2048 batches of two pairs identical to the ones above. The negative example in a pair is generated randomly for all models from the set of documents which are either labeled not relevant or not labeled for the query. Although different negative sampling mechanisms may impact final results \cite{lu2020neural}, the above simple negative sampling strategy achieves very good performance and has been successfully used in previous studies \cite{macavaney2019cedr,li2021keybld,pang2017deeprank}.
Gradient accumulation is employed every 8 steps to fit on a single GPU with 11GB memory like a RTX 2080ti GPU, simulating a batchsize with 16 training pairs, as done in \cite{macavaney2019cedr}. Automatic mixed precision \cite{micikevicius2018mixed} is used to speed up training. 
We adopt a validation mechanism with validation set to report each metric on the test set. That is to say, for each evaluation metric, we obtain the best performing model in the 10 epochs on the validation set, and use this model to obtain results on the test set for this metric. Each model is trained using Adam optimizer (the transformer layers are trained with a rate of $2*10^{-5}$ while the linear layer with a rate of $10^{-3}$). 
 
Results, for all models, are measured with P@1, P@5, P@10, P@20, MAP, NDCG@1, NDCG@5, NDCG@10, NDCG@20 and NDCG, NDCG and MAP being computed on all available documents for each query (the number of documents per query is 200 on Robust04 and GOV2 and varies from one query to the other on MQ2007 and MQ2008).
Each metric is calculated with \texttt{pytrec\_eval}\footnote{\url{https://github.com/cvangysel/pytrec_eval}} \cite{van2018pytrec_eval}, which is a wrapper of \texttt{trec\_eval}\footnote{\url{https://trec.nist.gov/trec_eval}}. Lastly, a paired t-test is used to assess whether differences are significant or not.

\subsection{Experimental results}

\label{compareApproach}
\begin{table*}[ht]
\centering
\caption{Results on {\it Robust04} dataset. Best results are in {\bf bold}. 
For KeyB(vBERT) models, a significant difference with BM25 is marked with a 'B', with DeepRank with a 'D', with Vanilla BERT with a 'V', with CEDR-KNRM with a 'C', and with Random select with an 'R'. For KeyB(PARADE) models, a significant difference with BM25 is marked with a 'B', with DeepRank with a 'D', with PARADE with a 'P', and with PARADE5 with a '5'.
A paired t-test ($p-value \le 0.05$) is used for measuring significance.}
\label{tab:rob}
\resizebox{\linewidth}{!}{
\begin{tabular}{lllllllllll}
\toprule
Model  & P@1 & P@5 & P@10 & P@20 & MAP& NDCG@1 & NDCG@5 & NDCG@10 & NDCG@20 & NDCG  \\ 
\hline\hline \multicolumn{11}{l}{\textit{\textbf{Baseline models}}} \\
\hline
BM25        & 0.5542  &0.5004  &0.4382 &0.3631 &0.2334&0.5080 &0.4741 &0.4485 &0.4240 & 0.4402  \\ 
DeepRank & 0.5663 & 0.4538 & 0.3907 & 0.3331 & 0.2145& 0.5081 & 0.4386 & 0.4051 & 0.3864 & 0.4272 \\
\hline\hline \multicolumn{11}{l}{\textit{\textbf{BERT based models}}} \\
\hline
Vanilla BERT   & 0.6067  & 0.5478  &0.4843 &0.4088&0.2510 &0.5706&0.5337&0.4945&0.4678&0.4553 \\

CEDR-KNRM & 0.6220 & 0.5542 & 0.4840 & 0.4097 & 0.2440 & 0.5878 & 0.5253 & 0.5093 & 0.4803 & 0.4600  \\

Random Select & 0.5983 & 0.5108  & 0.4730 &0.4059 & 0.2453&0.5482 &0.4856 &0.4880 &0.4688 &0.4540 \\
\hline
KeyB(vBERT)$_{TF-IDF}$  &0.6146  &0.5430$^{BDR}$  & 0.4963$^{BDR}$&0.4208$^{BDR}$&0.2628$^{BDVCR}$& 0.5764&0.5275$^{BDR}$&0.5099$^{BD}$&0.4884$^{BDVR}$&0.4684$^{BDVCR}$\\
KeyB(vBERT)$_{BM25}$  & 0.6468$^{BD}$ &0.5622$^{BDR}$   &0.4976$^{BDR}$&{\bf0.4241}$^{BDVR}$&0.2609$^{BDCR}$&0.6004$^{BD}$ &0.5512$^{BDR}$&0.5166$^{BDVR}$&0.4941$^{BDVR}$&0.4687$^{BDVCR}$ \\
KeyB(vBERT)$_{BinB}$   & {\bf0.6710}$^{BDR}$ &{\bf0.5661}$^{BDR}$  &{\bf0.5088}$^{BDVCR}$&{\bf0.4241}$^{BDVR}$&{\bf0.2722}$^{BDVCR}$&{\bf0.6289}$^{BDR}$&{\bf0.5554}$^{BDR}$&{\bf0.5249}$^{BDVR}$&{\bf0.4958}$^{BDVR}$&{\bf0.4768}$^{BDVCR}$ \\
\hline\hline \multicolumn{11}{l}{\textit{\textbf{PARADE based models}}} \\
\hline
PARADE  &0.6869 & 0.5686 & 0.5080 & 0.4309 & 0.2739& 0.6166 & 0.5510 & 0.5146 & 0.5017 & {0.4737} \\
PARADE5  & 0.6388  &0.5334  &0.4868 &0.4125 &0.2477&0.5905 &0.5215 &0.5055 &0.4761 & 0.4594  \\ 
\hline
KeyB(PARADE5)$_{TF-IDF}$ & {0.6790}$^{BD}$ & {0.5687}$^{BD5}$ & {0.5093}$^{BD5}$ & {0.4319}$^{BD5}$ &{0.2714}$^{BD5}$& {\bf0.6348}$^{BD}$ &{0.5467}$^{BD}$ &{0.5218}$^{BD}$ &{0.4989}$^{BD5}$ &0.4710$^{BD5}$ \\
KeyB(PARADE5)$_{BM25}$ & {\bf0.6871}$^{BD}$ & {\bf0.5768}$^{BD5}$ & {\bf0.5177}$^{BD5}$ & 0.4337$^{BD5}$ &0.2757$^{BD5}$& {0.6329}$^{BD}$ &{\bf0.5636}$^{BD5}$ &{\bf0.5304}$^{BD5}$ &{0.5040}$^{BD5}$ &0.4735$^{BD5}$ \\
KeyB(PARADE5)$_{BinB}$ & {0.6308}$^{B}$ & {0.5479}$^{BD}$ & {0.5057}$^{BD}$ & {0.4200}$^{BD}$ &{0.2629}$^{BDP5}$& {0.5885}$^{BD}$ &{0.5329}$^{BD}$ &{0.5283}$^{BD5}$ &{0.4967}$^{BD5}$ &0.4687$^{BD5}$ \\
KeyB(PARADE5)$_{BinB2}$ & {0.6427}$^{BD}$ & {0.5758}$^{BD5}$ & {0.5112}$^{BD5}$ & {\bf0.4378}$^{BD5}$& {\bf0.2779}$^{BD5}$ &{0.5965}$^{BD5}$ & {0.5625}$^{BD5}$ &0.5247$^{BD}$ & {\bf0.5058}$^{BD5}$&{\bf0.4778}$^{BD5}$ \\
 \bottomrule
\end{tabular}
}
\vspace{-0.3cm}
\end{table*}

\begin{table*}[ht]
\centering
\caption{Results on {\it GOV2} dataset. Best results are in {\bf bold}. Best results are in {\bf bold}.
For KeyB(vBERT) models, a significant difference with BM25 is marked with a 'B', with DeepRank with a 'D', with Vanilla BERT with a 'V', with CEDR-KNRM with a 'C', and with Random select with an 'R'. For KeyB(PARADE) models, a significant difference with BM25 is marked with a 'B', with DeepRank with a 'D', with PARADE with a 'P', and with PARADE5 with a '5'. 
 A paired t-test ($p-value \le 0.05$) is used for measuring significance.}
\label{tab:gov}
\resizebox{\linewidth}{!}{
\begin{tabular}{lllllllllll}
\toprule
Model  & P@1 & P@5 & P@10 & P@20 & MAP & NDCG@1 & NDCG@5 & NDCG@10 & NDCG@20 & NDCG  \\ 
\hline\hline \multicolumn{11}{l}{\textit{\textbf{Baseline models}}} \\
\hline
BM25        & 0.6510  &0.6054  &0.5792 &0.5362 &0.2331 &0.5034 &0.4904 &0.4867 &0.4774 & 0.4296  \\ 
DeepRank & 0.6453 & 0.5682 & 0.5143 & 0.4880 & 0.2151 & 0.4738 & 0.4363 & 0.4194 & 0.4170  & 0.4120  \\
\hline\hline \multicolumn{11}{l}{\textit{\textbf{BERT based models}}} \\
\hline
Vanilla BERT   & 0.6241 & 0.6068 & 0.5672 & 0.5475 &0.2321& 0.4531 &0.4954 &0.4837 &0.4764 &0.4279  \\

CEDR-KNRM &0.6239 &0.6133 &0.5886 &0.5556 &0.2375& 0.4929 & 0.4891 &0.4892 & 0.4769 &0.4315  \\
Random Select & 0.6839 & 0.6169  & 0.5984 &0.5640 & 0.2467&0.4995 &0.4811 &0.4955 &0.4853 &0.4358 \\
\hline

KeyB(vBERT)$_{TF-IDF}$  & 0.7122 & 0.6735$^{BDVCR}$  & 0.6446$^{BDVCR}$ &0.6123$^{BDVCR}$ & 0.2583$^{BDVCR}$&{\bf0.5574}$^{D}$ &0.5256$^{DR}$ &0.5340$^{BDVCR}$ &0.5269$^{BDCR}$ &0.4413$^{BDC}$  \\
KeyB(vBERT)$_{BM25}$  & {0.6634}  & 0.6524$^{D}$   & 0.6303$^{BDC}$ & 0.5997$^{BDVCR}$ &0.2643$^{BDVCR}$&{0.5171} &{0.5341}$^{DVR}$ &{0.5272}$^{BDVC}$ &0.5199$^{BDCR}$ &0.4447$^{BDVCR}$  \\

KeyB(vBERT)$_{BinB}$   & {\bf0.7651}$^{BDVC}$  & {\bf0.6937}$^{BDVCR}$  & {\bf0.6645}$^{BDVCR}$ & {\bf0.6125}$^{BDVCR}$ & {\bf0.2674}$^{BDVCR}$& {\bf0.5574}$^{D}$ & {\bf0.5414}$^{BDVCR}$ & {\bf0.5356}$^{BDVCR}$ & {\bf0.5295}$^{BDVCR}$ & {\bf0.4453}$^{BDVCR}$  \\
\hline\hline \multicolumn{11}{l}{\textit{\textbf{PARADE based models}}} \\
\hline
PARADE   & 0.7244 & 0.7016 & 0.6631 & 0.6133 & 0.2621& {\bf0.5930} & 0.5518 & {0.5562} & 0.5466 & 0.4484  \\ 
PARADE5       & 0.6906  &0.6429  &0.6246 &0.5707 &0.2462 &0.5463 &0.5327 &0.5161 &0.5053 & 0.4386 \\
\hline
KeyB(PARADE5)$_{TF-IDF}$ & {0.7386} & 0.6931$^{BD5}$ & 0.6605$^{BD5}$ & {0.6222}$^{BD5}$ &{0.2728}$^{BDP5}$ & 0.5569$^{D}$ &{0.5536}$^{BD}$ &0.5498$^{BD}$ &{0.5352}$^{BD5}$ &{0.4537}$^{BD5}$\\
KeyB(PARADE5)$_{BM25}$ & {\bf0.7720}$^{BD}$ & 0.6931$^{BD5}$ & 0.6528$^{BD}$ & 0.6397$^{BDP5}$ &{\bf0.2745}$^{BDP5}$ & {0.5806}$^{D}$ &{\bf0.5783}$^{BD5}$ &0.5529$^{BD5}$ &{\bf0.5624}$^{BD5}$ &{\bf0.4578}$^{BDP5}$\\
KeyB(PARADE5)$_{BinB}$ & {0.7055} & {\bf0.7196}$^{BD5}$ & {0.6563}$^{BD}$ & {0.6212}$^{BD5}$ &{0.2680}$^{BD5}$& {0.5640}$^{D}$ &{0.5660}$^{BD}$ &{0.5451}$^{BD}$ &{0.5379}$^{BD}$ &0.4495$^{BD}$ \\
KeyB(PARADE5)$_{BinB2}$ & {0.7253} & {0.7034}$^{BD5}$ & {\bf0.6771}$^{BD5}$ & {\bf0.6407}$^{BDP5}$ &{0.2733}$^{BDP5}$& {0.5706}$^{D}$ &{0.5512}$^{BD}$ &{\bf0.5676}$^{BD5}$ &{0.5591}$^{BD5}$ &0.4554$^{BD5}$ \\
\bottomrule
\end{tabular}
}
\end{table*}

\begin{table*}[ht]
\centering
\caption{Results on {\it MQ2007} dataset. DeepRank* represents the results from the original paper. Best results are in {\bf bold}.
For KeyB(vBERT) models, a significant difference with BM25 is marked with a 'B', with DeepRank with a 'D', with Vanilla BERT with a 'V', with CEDR-KNRM with a 'C', and with Random select with an 'R'. For KeyB(PARADE) models, a significant difference with BM25 is marked with a 'B', with DeepRank with a 'D', with PARADE with a 'P', and with PARADE5 with a '5'.
 A paired t-test ($p-value \le 0.05$) is used for measuring significance.}
\label{tab:mq2007}
\resizebox{\linewidth}{!}{
\begin{tabular}{lllllllllll}
\toprule
Model  & P@1 & P@5 & P@10 &P@20 & MAP & NDCG@1 & NDCG@5 & NDCG@10 & NDCG@20 & NDCG\\ 
\hline\hline \multicolumn{11}{l}{\textit{\textbf{Baseline models}}} \\
\hline
BM25        &0.4186 & 0.3969 & 0.3757 & 0.3391& 0.4527& 0.3712 & 0.3954 & 0.4309 &0.4962 & 0.5933  \\ 
DeepRank       &  0.4444 & 0.4201 & 0.3898 & 0.3473 & 0.4596& 0.3942 & 0.4168 & 0.4468 & 0.5088 & 0.6012  \\
DeepRank*       &  0.508 & 0.452 & 0.412 &- & 0.497& 0.441 & 0.457 & 0.482  & -& -  \\
\hline\hline \multicolumn{11}{l}{\textit{\textbf{BERT based models}}} \\
\hline
Vanilla BERT   & 0.5266 & 0.4741 &0.4257 & 0.3606 &0.5073&0.4708 &0.4808 &0.5070 &0.5620&0.6379  \\
CEDR-KNRM  & 0.5284 & 0.4768 & 0.4233 & 0.3601& 0.5066& 0.4814 & 0.4874 & 0.5084 &0.5601& 0.6380  \\
Random Select  & 0.5343 & 0.4768 &0.4347 &0.3656&0.5207&0.4808 &0.4980 &0.5224 &0.5775&0.6499   \\
\hline
KeyB(vBERT)$_{TF-IDF}$  & 0.5425$^{BD}$ & 0.4926$^{BDVCR}$ &0.4465$^{BDVCR}$ &0.3702$^{BDVCR}$ &0.5323$^{BDVCR}$&0.4917$^{BD}$ &0.5043$^{BDVC}$ &0.5342$^{BDVCR}$ &0.5864$^{BDVCR}$ &0.6551$^{BDVC}$   \\
KeyB(vBERT)$_{BM25}$  & 0.5526$^{BDVC}$  & 0.4946$^{BDVCR}$  & 0.4408$^{BDVC}$ &0.3705$^{BDVCR}$&0.5305$^{BDVCR}$&0.4933$^{BDV}$&0.5061$^{BDVC}$&0.5339$^{BDVCR}$&0.5824$^{BDVC}$&0.6528$^{BDVC}$  \\
KeyB(vBERT)$_{BinB}$   & {\bf0.5597}$^{BDVCR}$ & {\bf0.4971}$^{BDVCR}$ &{\bf0.4503}$^{BDVCR}$ &{\bf0.3759}$^{BDVCR}$ &{\bf0.5457}$^{BDVCR}$&{\bf0.5133}$^{BDVCR}$ &{\bf0.5134}$^{BDVCR}$ &{\bf0.5496}$^{BDVCR}$ &{\bf0.5969}$^{BDVCR}$&{\bf0.6627}$^{BDVCR}$  \\ 
\hline\hline \multicolumn{11}{l}{\textit{\textbf{PARADE based models}}} \\
\hline
PARADE & 0.5474 & 0.5009 & {0.4486} & 0.3747 & 0.5418& 0.5054 & 0.5255 & 0.5499 &0.5950& 0.6599   \\
PARADE5       & 0.5686  &0.4824  &0.4370 & 0.3714 &0.5291&{0.5174} &0.5142 &0.5356 & 0.5851& 0.6538   \\ 
\hline
KeyB(PARADE5)$_{TF-IDF}$ & {0.5721}$^{BDP}$ & {0.5034}$^{BD5}$ & {\bf0.4491}$^{BD5}$ & {0.3737}$^{BD}$& {0.5477}$^{BD5}$&0.5198$^{BD}$ & {0.5221}$^{BD}$ &{0.5488}$^{BD5}$ & {0.5998}$^{BD5}$& {0.6645}$^{BD5}$ \\
KeyB(PARADE5)$_{BM25}$ & {\bf0.5769}$^{BDP}$ & {0.5063}$^{BD5}$ & {0.4486}$^{BD5}$ & {0.3748}$^{BD5}$& {\bf0.5494}$^{BDP5}$&0.5151$^{BD}$ & {0.5261}$^{BD5}$ &{\bf0.5530}$^{BD5}$ & {\bf0.6021}$^{BD5}$& {\bf0.6664}$^{BDP5}$ \\
KeyB(PARADE5)$_{BinB}$ & {0.5580}$^{BD}$ & {\bf0.5079}$^{BD5}$ & {0.4487}$^{BD5}$ & {0.3740}$^{BD}$& {0.5427}$^{BD5}$&0.5054$^{BD}$ & {0.5200}$^{BD}$ &{0.5493}$^{BD5}$ & {0.5978}$^{BD5}$& {0.6623}$^{BD5}$ \\
KeyB(PARADE5)$_{BinB2}$ & {0.5709}$^{BDP}$ & {0.5066}$^{BD5}$ & {0.4488}$^{BD5}$ & {\bf0.3766}$^{BD5}$& {0.5461}$^{BD5}$&{\bf0.5213}$^{BD}$ & {\bf0.5266}$^{BD5}$ &{0.5513}$^{BD5}$ & {0.6005}$^{BD5}$& {0.6650}$^{BD5}$ \\
\bottomrule

\end{tabular}
}
\end{table*}

\begin{table*}[ht]
\centering
\caption{Results on {\it MQ2008} dataset. DeepRank* represents the results from the original paper. Best results are in {\bf bold}. 
For KeyB(vBERT) models, a significant difference with BM25 is marked with a 'B', with DeepRank with a 'D', with Vanilla BERT with a 'V', with CEDR-KNRM with a 'C', and with Random select with an 'R'. For KeyB(PARADE) models, a significant difference with BM25 is marked with a 'B', with DeepRank with a 'D', with PARADE with a 'P', and with PARADE5 with a '5'.
A paired t-test ($p-value \le 0.05$) is used for measuring significance.}
\label{tab:mq2008}
\resizebox{\linewidth}{!}{
\begin{tabular}{lllllllllll}
\toprule
Model  & P@1 & P@5 & P@10 &P@20 & MAP & NDCG@1 & NDCG@5 & NDCG@10 & NDCG@20 & NDCG  \\ 
\hline\hline \multicolumn{11}{l}{\textit{\textbf{Baseline models}}} \\
\hline
BM25        & 0.3816   & 0.3316 & 0.2411 & 0.1515& 0.4538& 0.3297 & 0.4376& 0.4841& 0.5086& 0.5243 \\ 
DeepRank & 0.3992 & 0.2816 & 0.1920 &0.1150 & 0.4356& 0.3641 & 0.4373 & 0.4672 & 0.4878 & 0.4917  \\
DeepRank* & 0.482 & 0.359 & 0.252 & - & 0.498& 0.406 & 0.496 & - &- & -  \\
\hline\hline \multicolumn{11}{l}{\textit{\textbf{BERT based models}}} \\
\hline
Vanilla BERT   & 0.5063 & 0.3650 &0.2560 & 0.1566 & 0.5230&0.4508 &0.5165 &0.5489 & 0.5697 & 0.5810  \\

CEDR-KNRM  & 0.5050 & 0.3678 & 0.2561 & 0.1569 & 0.5220& 0.4515 & 0.5151 & 0.5488 & 0.5674 & 0.5794  \\ 
Random Select  & 0.5000 & 0.3663 &0.2574 & 0.1579& 0.5196&0.4387 &0.5096 &0.5427 & 0.5611 & 0.5729   \\
\hline
KeyB(vBERT)$_{TF-IDF}$  & 0.5166$^{BD}$  & {\bf0.3862}$^{BDVCR}$  & 0.2597$^{D}$ &0.1580$^{D}$ &0.5318$^{BD}$&0.4649$^{BD}$ & 0.5330$^{BDVCR}$ &0.5596$^{BDR}$ &0.5755$^{BDR}$ &0.5869$^{BDR}$ \\
KeyB(vBERT)$_{BM25}$ & 0.5165$^{BD}$ & 0.3760$^{BDVR}$ & 0.2579$^{D}$ & 0.1582$^{D}$&  0.5350$^{BDR}$& 0.4629$^{BD}$& 0.5317$^{BDVCR}$ & 0.5609$^{BDVCR}$&{\bf0.5788}$^{BDCR}$ & {\bf0.5891}$^{BDR}$ \\
KeyB(vBERT)$_{BinB}$   & {\bf0.5254}$^{BD}$& 0.3819$^{BDVCR}$& {\bf0.2624}$^{DVCR}$& {\bf0.1589}$^{D}$ & {\bf0.5425}$^{BDVCR}$&{\bf0.4661}$^{BDR}$ &{\bf0.5382}$^{BDVCR}$ & {\bf0.5616}$^{BDVCR}$& {\bf0.5788}$^{BDCR}$ & {\bf0.5891}$^{BDR}$   \\
\hline\hline \multicolumn{11}{l}{\textit{\textbf{PARADE based models}}} \\
\hline
PARADE & 0.5089 & 0.3811 & 0.2617 & 0.1590& 0.5375& 0.4502 & 0.5321 & {0.5656} & 0.5799& 0.5867   \\
PARADE5       & 0.4999  &0.3763  &0.2578 & 0.1574 &0.5254 &0.4514 &0.5226 &0.5523 & 0.5716 & 0.5821  \\ 
\hline
KeyB(PARADE5)$_{TF-IDF}$ & {\bf0.5369}$^{BDP5}$ & {0.3829}$^{BD}$ & {0.2621}$^{D5}$ & {0.1585}$^{D}$& {\bf0.5436}$^{BD5}$ &{\bf0.4859}$^{BDP5}$ & {\bf0.5390}$^{BD5}$ &\bf0.5728$^{BD5}$ & {\bf0.5851}$^{BD5}$&{0.5942}$^{BD5}$ \\
KeyB(PARADE5)$_{BM25}$ & {0.5267}$^{BD5}$ & {0.3831}$^{BD}$ & {0.2635}$^{D5}$ & {\bf0.1592}$^{D5}$& {0.5409}$^{BD5}$ &{0.4744}$^{BDP}$ & {0.5373}$^{BD5}$ &0.5646$^{BD5}$ & {0.5812}$^{BD}$&{0.5907}$^{BD5}$ \\
KeyB(PARADE5)$_{BinB}$ & {0.5229}$^{BD}$ & {\bf0.3888}$^{BDP5}$ & {0.2634}$^{D5}$ & {\bf0.1592}$^{D}$& {0.5428}$^{BD5}$ &{0.4668}$^{BD}$ & {0.5354}$^{BD5}$ &0.5702$^{BD5}$ & {0.5835}$^{BD5}$&{0.5914}$^{BD}$ \\
KeyB(PARADE5)$_{BinB2}$ & {0.5242}$^{BD}$ & {0.3847}$^{BD5}$ & {\bf0.2639}$^{D5}$ & {0.1583}$^{D}$& {0.5431}$^{BD5}$ &{0.4789}$^{BDP5}$ & {0.5373}$^{BD5}$ &0.5689$^{BD5}$ & {0.5837}$^{BD5}$&{\bf0.5952}$^{BD5}$ \\

\bottomrule
\end{tabular}
}
\end{table*}

The results obtained on all the four collections are displayed in Tables \ref{tab:rob} to \ref{tab:mq2008}\footnote{As pointed out in e.g. \url{https://github.com/Georgetown-IR-Lab/cedr/issues/22}, the results obtained for CEDR-KNRM with the code provided by the authors differ from the ones reported in the original paper. We have also observed this in our experiments. The same holds for DeepRank; in that case however the original paper provides results for most of the metrics we have retained on MQ2007 and MQ2008. We have thus reported the original results in Tables \ref{tab:mq2007} and \ref{tab:mq2008}, under the name DeepRank$^{*}$. These results do not change our conclusions.}. We first analyze the results of the KeyB(vBERT) models, prior to analyze the ones of KeyB(PARADE5)$_{BM25}$ and compare them. 

\subsubsection{Improving Vanilla BERT model with selected key blocks}

We propose to analyze the experimental results by answering several research questions.
\newcommand{\RQone}{\begin{itemize}
    \item[\textbf{RQ1}] How effective are KeyB(vBERT) models compared to baseline models (BM25, DeepRank)?
\end{itemize}}
\RQone

\noindent The first conclusion we draw from Tables \ref{tab:rob} to \ref{tab:mq2008} is that all KeyB(vBERT) models outperform both baseline models on all collections, for all metrics, by a large margin.
Furthermore, on all collections, KeyB(vBERT) models significantly outperform most of metrics.

\newcommand{\RQtwo}{\begin{itemize}
    \item[\textbf{RQ2}] How effective are KeyB(vBERT) models compared to standard BERT based models (Vanilla BERT, CEDR-KNRM)?
\end{itemize}}
\RQtwo

\noindent As one can see, 
KeyB(vBERT)$_{BM25}$ and KeyB(vBERT)$_{BinB}$ models outperform standard BERT models (Vanilla BERT and CEDR-KNRM) on all collections and for all metrics. KeyB(vBERT)$_{TF-IDF}$ outperforms standard BERT models (Vanilla BERT and CEDR-KNRM) for all metrics on all collections except Robust04, for which it yields lower results than Vanilla BERT on P@5, NDCG@5, and lower result than CEDR-KNRM on P@1, P@5 and NDCG@1.
In addition, the best KeyB(vBERT) model (on each metric respectively) significantly improves the Vanilla BERT model on 6 metrics out of 10 on Robust04, on 9 metrics out of 10 on GOV2, on all metrics on MQ2007 and on 5 metrics out of 10 on MQ2008; it is furthermore significantly better than CEDR-KNRM on 2 metrics on Robust04, on 9 metrics out of 10 on GOV2, on all metrics on MQ2007, and on 6 Metrics out of 10 on MQ2008.

\newcommand{\RQthree}{\begin{itemize}
    \item[\textbf{RQ3}] Is it important to accurately select blocks?
\end{itemize}}
\RQthree

\noindent We are interested here in assessing whether it is important to accurately select blocks or not. For this, we compare the results obtained by the different KeyB(vBERT) models with the ones obtained by the Random Select strategy which amounts to randomly selecting blocks. As one can also note, all KeyB(vBERT) models outperform the Random Select strategy on all collections, for all metrics.
 Furthermore, the best KeyB(vBERT) model is significantly better than Random Select on 10 metrics out of 10 on Robust04, on 8 metrics out of 10 on GOV2, on 10 metrics out of 10 on MQ2007, and on 8 Metrics out of 10 on MQ2008.

The above analysis shows that the KeyB(vBERT) models should be preferred over all baseline and standard BERT-based IR models. We now turn to the comparison of KeyB(vBERT) models.

\newcommand{\RQfour}{\begin{itemize}
    \item[\textbf{RQ4}] What are the differences between the different KeyB(vBERT) models?
\end{itemize}}
\RQfour

\noindent On Robust04, KeyB(vBERT)$_{BinB}$ is the best model on 10 metrics and KeyB(vBERT)$_{BM25}$ on 1 metric. We further conduct significant tests between KeyB(vBERT) models. KeyB(vBERT)$_{BinB}$ is significantly better than KeyB(vBERT)$_{BM25}$ on MAP and NDCG while shows no significant difference than KeyB(vBERT)$_{TF-IDF}$. On GOV2, KeyB(vBERT)$_{BinB}$ is the best model on 10 metrics and KeyB(vBERT)$_{TF-IDF}$ on 1 metric. KeyB(vBERT)$_{BinB}$ is significantly better than KeyB(vBERT)$_{BM25}$ on p@1 and p@10 while shows no significant difference than KeyB(vBERT)$_{TF-IDF}$. On MQ2007, KeyB(vBERT)$_{BinB}$ is the best model over all metrics and significantly outperforms KeyB(vBERT)$_{TF-IDF}$ on MAP, P@20, NDCG, NDCG@1, NDCG@10 and NDCG@20,  significantly outperforms KeyB(vBERT)$_{BM25}$ on MAP, P@10, P@20, NDCG, NDCG@10 and NDCG@20. On MQ2008, KeyB(vBERT)$_{BinB}$ is the best model on all metrics but P@5, KeyB(vBERT)$_{TF-IDF}$ on 1 metric and KeyB(vBERT)$_{BM25}$ on 2 metrics.
The difference between all three models is however not really significant as KeyB(vBERT)$_{BinB}$ significantly outperforms KeyB(vBERT)$_{TF-IDF}$ on only MAP and KeyB(vBERT)$_{BM25}$ on P@10.

From this analysis, one can see that the model  KeyB(vBERT)$_{BinB}$ is either significantly better or on a par with KeyB(vBERT)$_{TF-IDF}$ and KeyB(vBERT)$_{BM25}$. This justifies the use of a learning mechanism to select blocks. This said, even a simple approach to select blocks as the one implemented in KeyB(vBERT)$_{BM25}$ can yield good results on collections such as Robust04 and MQ2008. We now turn to the PARADE models.

\subsubsection{Improving PARADE with selected passages}

As mentioned before, PARADE is the original PARADE model with 16 passages corresponding to the first and last passages, and 14 randomly selected passages in between, PARADE5 is another variant with only 5 passages corresponding to the first and last passages, and 3 randomly selected passages in between, and KeyB(PARADE5) models are the PARADE models with only 5 passages selected with BM25, TF-IDF or learning based approaches. We propose to analyze the experimental results by answering several research questions.

\newcommand{\RQfive}{\begin{itemize}
  \item[\textbf{RQ5}] How effective are KeyB(PARADE5) models compared to baseline models (BM25, DeepRank)?
\end{itemize}}
\RQfive

\noindent From Tables \ref{tab:rob} to \ref{tab:mq2008}, one can see that KeyB(PARADE5) models outperform both baselines on all collections, the difference being significant for all metrics and all on collections but P@1 on GOV2.

\newcommand{\RQsix}{\begin{itemize}
    \item[\textbf{RQ6}] How effective are KeyB(PARADE5) models compared to PARADE and PARADE5?
\end{itemize}}
\RQsix

\noindent As one can note, on all collections, KeyB(PARADE5)$_{BM25}$, KeyB(PARADE5)$_{BM25}$ and KeyB(PARADE5)$_{BinB2}$ obtain better average results or on a par with PARADE. For example, comparing with the original PARADE model, on Robust04, KeyB(PARADE5)$_{BM25}$ outperforms PARADE on 9 metrics out of 10, even though the difference is never significant. On GOV2, KeyB(PARADE5)$_{BinB2}$ outperforms PARADE on 8 metrics out of 10 and is significantly better on 2 metrics. On MQ2007, KeyB(PARADE5)$_{BM25}$ outperforms PARADE on 9 metrics out of 10 and is significantly better on 3 metrics. On MQ2008, KeyB(PARADE5)$_{TF-IDF}$ outperforms PARADE on 9 metrics out of 10 and is significantly better on 2 metrics.

Comparing with PARADE5, KeyB(PARADE5)$_{BM25}$, KeyB(PARADE5)$_{BM25}$ and KeyB(PARADE5)$_{BinB2}$ obtain better average results on all collections and metrics, except KeyB(PARADE5)$_{BM25}$ on NDCG@1 on MQ2007 (0.5151 vs 0.5174). More precisely, on Robust04, KeyB(PARADE5)$_{BM25}$ and KeyB(PARADE5)$_{BinB2}$ are significantly better than PARADE5 on 8 metrics. On GOV2, KeyB(PARADE5)$_{BM25}$ and KeyB(PARADE5)$_{BinB2}$ are significantly better than PARADE5 on 7 metrics. On MQ2007, KeyB(PARADE5)$_{TF-IDF}$ and KeyB(PARADE5)$_{BM25}$ are significantly better than PARADE5 on 6 and 8 metrics respectively. On MQ2008, KeyB(PARADE5)$_{TF-IDF}$ and KeyB(PARADE5)$_{BinB2}$ are significantly better than PARADE5 on 8 metrics.

  The model KeyB(PARADE5)$_{BinB}$, which reuses the BERT and feed-forward neural networks in PARADE for selecting passages, is however sometimes less effective than the other KeyB(PARADE5) models. On Robust04, it is below the other three KeyB(PARADE5) models as well as below PARADE. On GOV2, KeyB(PARADE5)$_{BinB}$ is higher than PARADE on 5 metrics and lower on 5 metrics. On MQ2007 and MQ2007, KeyB(PARADE5)$_{BinB}$ is mostly better than PARADE. Comparing with PARADE5, KeyB(PARADE5)$_{BinB}$ obtains almost always better results, especially on MQ2007 and MQ2008, being significantly better on 6 and 5 metrics respectively. This shows that, despite its mitigated results on some metrics and collections, KeyB(PARADE5)$_{BinB}$ is still a powerful approach that obtains several best results on different metrics.

\newcommand{\RQseven}{\begin{itemize}
  \item[\textbf{RQ7}] What are the differences between the different KeyB(PARADE5) models?
\end{itemize}}
\RQseven

\noindent As one can note, the best results on each metric is somehow distributed on the different KeyB(PARADE5) models. On Robust04, KeyB(PARADE5)$_{BM25}$ obtains 5 best results, KeyB(PARADE5)$_{BinB2}$ 4 and KeyB(PARADE5)$_{TF-IDF}$ 1. On GOV2, KeyB(PARADE5)$_{BM25}$ obtains 5 best results, KeyB(PARADE5)$_{BinB2}$ 3 and KeyB(PARADE5)$_{BinB}$ 1. On MQ2007, KeyB(PARADE5)$_{BM25}$ obtains 5 best results, KeyB(PARADE5)$_{BinB2}$ 3 and the other two models 1 each. On MQ2008, KeyB(PARADE5)$_{TF-IDF}$ obtains 6 best results, KeyB(PARADE5)$_{BM25}$ 1 and KeyB(PARADE5)$_{BinB}$ and KeyB(PARADE5)$_{BinB}$ 2 each. Besides, as discussed above, although KeyB(PARADE5)$_{BinB}$ does not perform well on Robust04, it is still competitive with PARADE5 on this collection and performs well on the other collections. KeyB(PARADE5)$_{BinB2}$ tends however to be more stable across the collections and metrics.

Overall, the PARADE variants we have introduced in general significantly outperform the PARADE5 model and are either on a par or significantly outperform the original PARADE model. This is all the more remarkable that these models use three times less passages than the original PARADE model and require less memory while being faster, as illustrated below. Lastly, the best KeyB(PARADE5) model tends to be slightly better than the best KeyB(vBERT) model, on all collections and almost all metrics, even though the difference is in general small. Their latency is however not the same (see below).

\subsection{Memory usage}
\label{memoryCompaSec}

\begin{figure*}[tb]
    \centering
    \begin{subfigure}[t]{0.3\textwidth}
        \centering
        \includegraphics[height=1.8in]{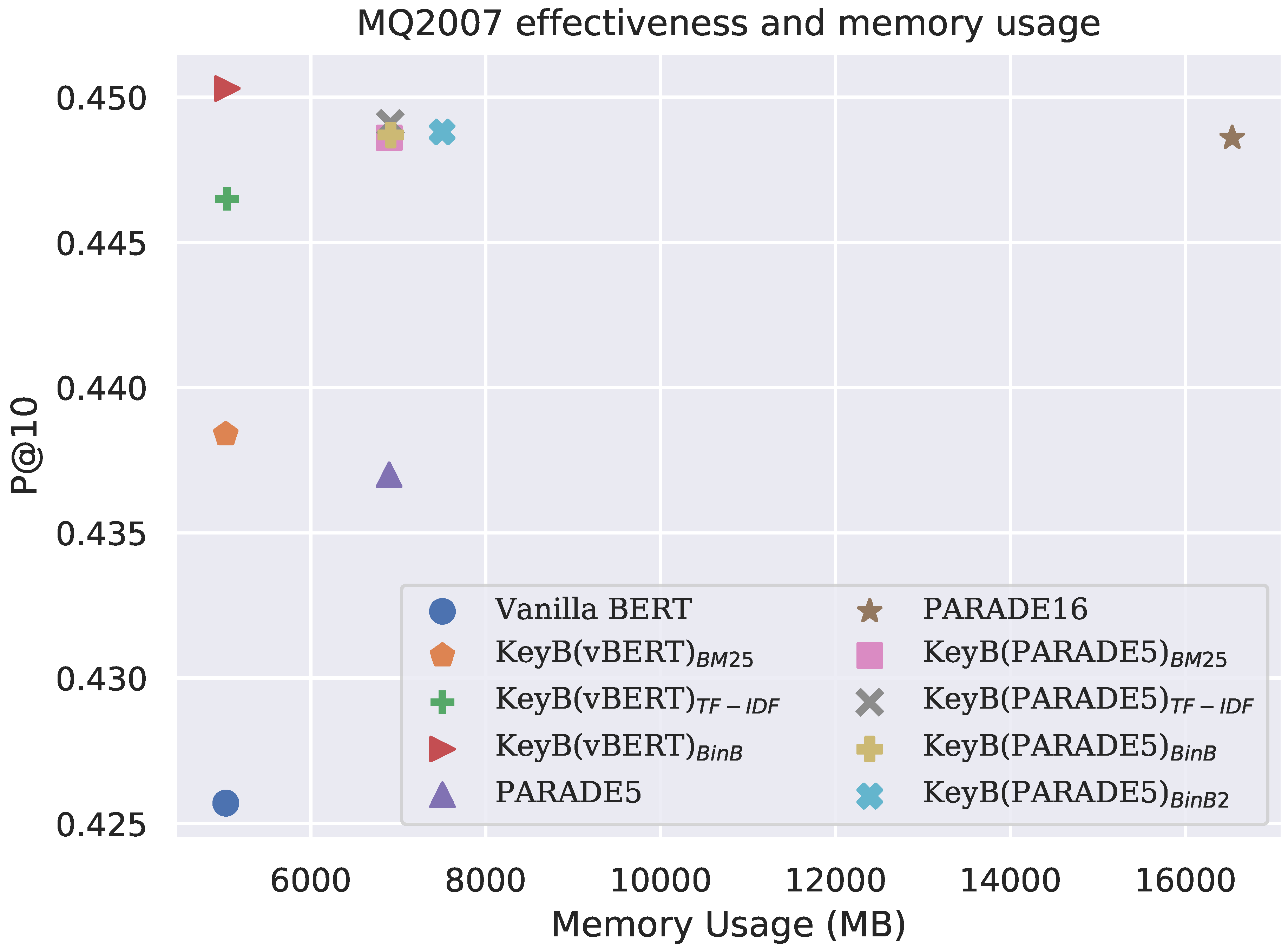}
        \caption{P@10 performance}
    \end{subfigure}%
    \hspace{0.8in}
    \begin{subfigure}[t]{0.3\textwidth}
        \centering
        \includegraphics[height=1.8in]{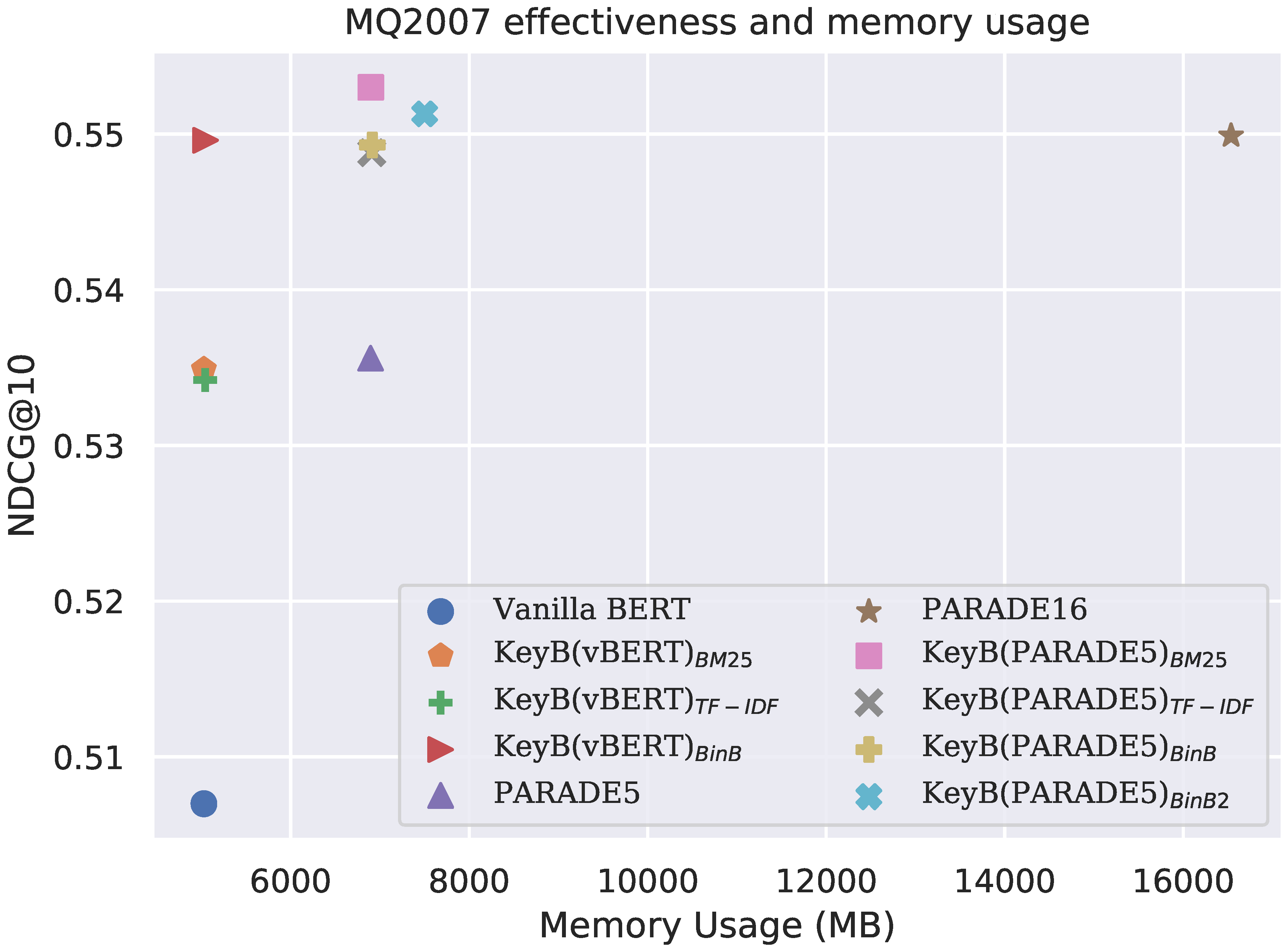}
        \caption{NDCG@10 performance}
    \end{subfigure} %
    \caption{GPU memory usage and effectiveness comparisons, automatic mixed precision is used for all models which would reduce memory usage. Top left models show better performance.}
    \label{fig.memoryMq2007}
\end{figure*}

The memory usage of all models are similar across datasets. We thus only report here the memory usage of different models on MQ2007 as this dataset contains more queries and requires longer training than Robust04 and GOV2. The memory usage corresponds to the GPU consumption for training a given model. We remind the reader that a training batch contains two pairs consisting of four queries and four documents. The results obtained are shown in Figure~\ref{fig.memoryMq2007} on two official LETOR metrics \cite{qin2013introducing}.

The best models in terms of accuracy (as measured by either P@10 or NDCG@10) and memory usage are located in the top left corner: they use less memory and  achieve higher results. As one can note, KeyB(vBERT)$_{BinB}$ and KeyB(PARADE5) models are located in this area. They need less GPU memory while achieving similar or higher results than PARADE on the two metrics. Furthermore, KeyB(PARADE5)$_{TF-IDF}$, KeyB(PARADE5)$_{BM25}$ and KeyB(PARADE5)$_{BinB}$ uses the same amount of memory as PARADE5 but are better on both metrics. KeyB(PARADE5)$_{BinB2}$ uses slightly more memory than PARADE5 but is also better on both metrics.

\subsection{Ranking speed}
\label{speedCompSec}

We measure here the speed of ranking of the different models on two sets of queries: all queries from one test fold of Robust04, each with 200 documents, and a randomly selected subset of 100 queries from MQ2007, again from one test fold, each with 40 documents. Note that the documents in MQ2007 are on average longer than the ones in Robust04. Latency results, as well as the average time for processing a query (in seconds) and a document (in milliseconds) on a RTX 6000 GPU are reported in Table~\ref{tab:latency} for Robust04 and Table~\ref{tab:latencyMq7} for MQ2007. The passage splitting time is not counted as this step can be performed offline. 

As one can see on both tables, the three fastest models are Vanilla BERT, KeyB(vBERT)$_{TF-IDF}$ and KeyB(vBERT)$_{BM25}$, the latter two being only slightly slower than the former one. Furthermore, on both collections, the KeyB(PARADE5)$_{TF-IDF}$ and KeyB(PARADE5)$_{BM25}$ are faster than PARADE. They are also faster than PARADE5 on Robust04 and only slightly slower than PARADE5 on MQ2007. These two variants, TF-IDF and BM25, because of their performance, their memory usage and their speed, represent strong alternatives to the original Vanilla BERT and PARADE models.

Regarding the models based on learning the selection block method, if their performance is higher than the one of other models, their latency is also higher: KeyB(vBERT)$_{BinB}$, KeyB(PARADE5)$_{BinB}$ and KeyB(PARADE5)$_{BinB2}$ are, at best, 10 times slower than the KeyB(vBERT)$_{TF-IDF}$ and KeyB(vBERT)$_{BM25}$ models on both collections. Their current latency may prevent their use in a commercial system. This said, there are several paths that one can follow to make them faster, including a two-stage approach at the block level, using a fast model as BM25 for filtering out less relevant blocks and using the more complex models on the remaining blocks.

\begin{table}[t]
  \centering
  \caption{Reranking latencies (seconds) on {\it Robust04} test set for one folder (50 queries each with 200 documents). }
  \label{tab:latency}
  \scalebox{0.9}{
  \begin{tabular}{cccc}
    \toprule
    Model  & Latency & Seconds/query & Milliseconds/doc \\ \hline
    Vanilla BERT  & 16.784 &  0.336 & 1.678 \\
    \hline
    KeyB(vBERT)$_{TF-IDF}$  & 18.475 & 0.370 & 1.848\\
    KeyB(vBERT)$_{BM25}$     & 19.679 & 0.394 & 1.970\\
    KeyB(vBERT)$_{BinB}$     &  178.339 & 3.567& 17.834\\
    \hline
    PARADE     &  75.601&1.512&7.560 \\
    PARADE5     & 55.661 & 1.113&5.566\\
    \hline
    KeyB(PARADE5)$_{TF-IDF}$     & 47.210 & 0.944&4.721\\
    KeyB(PARADE5)$_{BM25}$     & 47.781 &0.956&4.778 \\
    KeyB(PARADE5)$_{BinB}$     & 135.499 & 2.710&13.550\\
    KeyB(PARADE5)$_{BinB2}$     & 175.562 & 3.511&17.556\\
    \bottomrule
    \end{tabular}
  }
  \end{table}

  \begin{table}[t]
    \centering
    \caption{Ranking latencies (seconds) on {\it MQ2007} test set for 100 queries each with 40 documents. }
    \label{tab:latencyMq7}
    \scalebox{0.9}{
    \begin{tabular}{cccc}
    \toprule
    Model  & Latency & Seconds/query & Milliseconds/doc\\ \hline
    Vanilla BERT  & 6.962 & 0.070  &1.741 \\
    \hline
    KeyB(vBERT)$_{TF-IDF}$  & 11.598 & 0.116& 2.900\\
    KeyB(vBERT)$_{BM25}$     & 13.742 & 0.137 &3.436 \\
    KeyB(vBERT)$_{BinB}$     &  211.648 & 2.12& 52.912\\
    \hline
    PARADE     &  27.909 & 0.279& 6.977\\
    PARADE5     & 18.701 & 0.187&4.675\\
    \hline
    KeyB(PARADE5)$_{TF-IDF}$     & 26.894 & 0.269 &6.724\\
    KeyB(PARADE5)$_{BM25}$     & 24.320  & 0.243 &6.080\\
    KeyB(PARADE5)$_{BinB}$     & 242.309 & 2.423&60.577\\
    KeyB(PARADE5)$_{BinB2}$     & 336.789 & 3.368&84.197\\
    \bottomrule
    \end{tabular}
    }
    \end{table}

\subsection{Analysis of the position of selected blocks}

\begin{figure}[!t] \centering \includegraphics[width=0.55\textwidth]{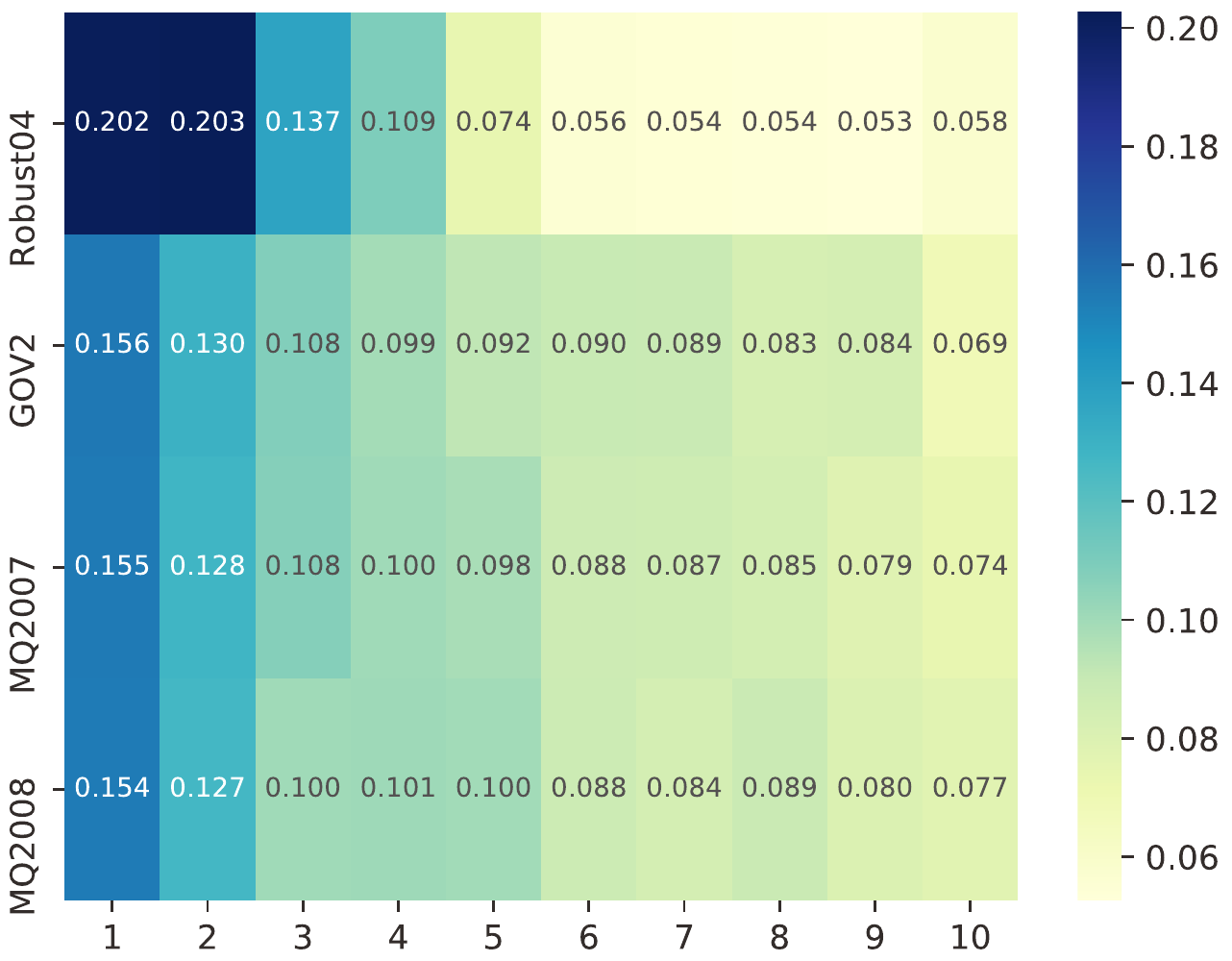}
  \caption{The probabilities of top 8 block appearing locations in KeyB(vBERT)$_{BM25}$.}
  \label{fig.stateBm25Blocks} 
  \end{figure}

\begin{figure}[!t] \centering \includegraphics[width=0.55\textwidth]{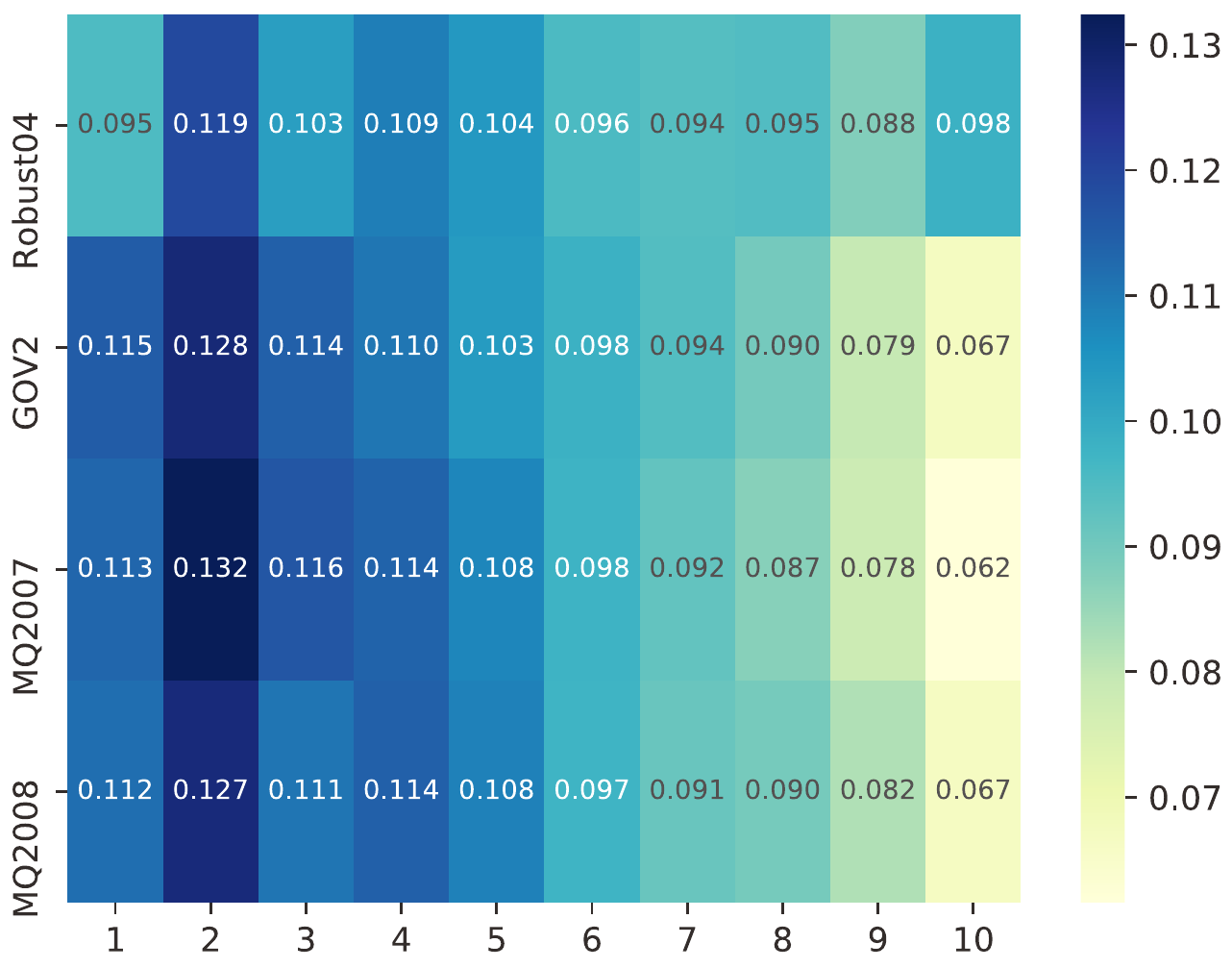}
  \caption{The probabilities of top 8 block appearing locations in KeyB(vBERT)$_{BinB}$.}
  \label{fig.stateBinBlocks} 
  \end{figure}

We are finally interested here in analyzing the positions at which blocks are selected. To do so, we retained all documents containing at least 15 blocks and looked at which position the top eight scoring blocks occur for two models with different selection strategies, namely KeyB(vBERT)$_{BM25}$ and KeyB(vBERT)$_{BinB}$. The position is computed as in Section~\ref{sec:analysis}. The results obtained are displayed in Figure~\ref{fig.stateBm25Blocks} for KeyB(vBERT)$_{BM25}$ and in Figure~\ref{fig.stateBinBlocks} for KeyB(vBERT)$_{BinB}$. In both Figures, a heat map is used to represent the probability the blocks selected occupy a particular position.

As one can observe, both models are more likely to select blocks at the beginning of a document, this tendency being more marked in KeyB(vBERT)$_{BM25}$. Interestingly, KeyB(vBERT)$_{BinB}$ is more likely to select blocks in the second position (and even in the third position) than in the first position. This said, both models also have a non null probability to select blocks at later positions. For example, in MQ2008, the possibility of selecting blocks in the last five positions, \textit{i.e.}, in the second half of a document, amounts to 41.8\% for KeyB(vBERT)$_{BM25}$ and to 42.7\% for KeyB(vBERT)$_{BinB}$. The situation is similar on MQ2007 and GOV2, as well as on Robust04 for KeyB(vBERT)$_{BinB}$. These results, in line with the analysis conducted in Section~\ref{sec:analysis}, show the capacity of the proposed models to rely on blocks at different positions in the document; the good performance of these models further shows that the blocks selected tend to contain relevant information.

\begin{figure}[!t] \centering \includegraphics[width=0.95\textwidth]{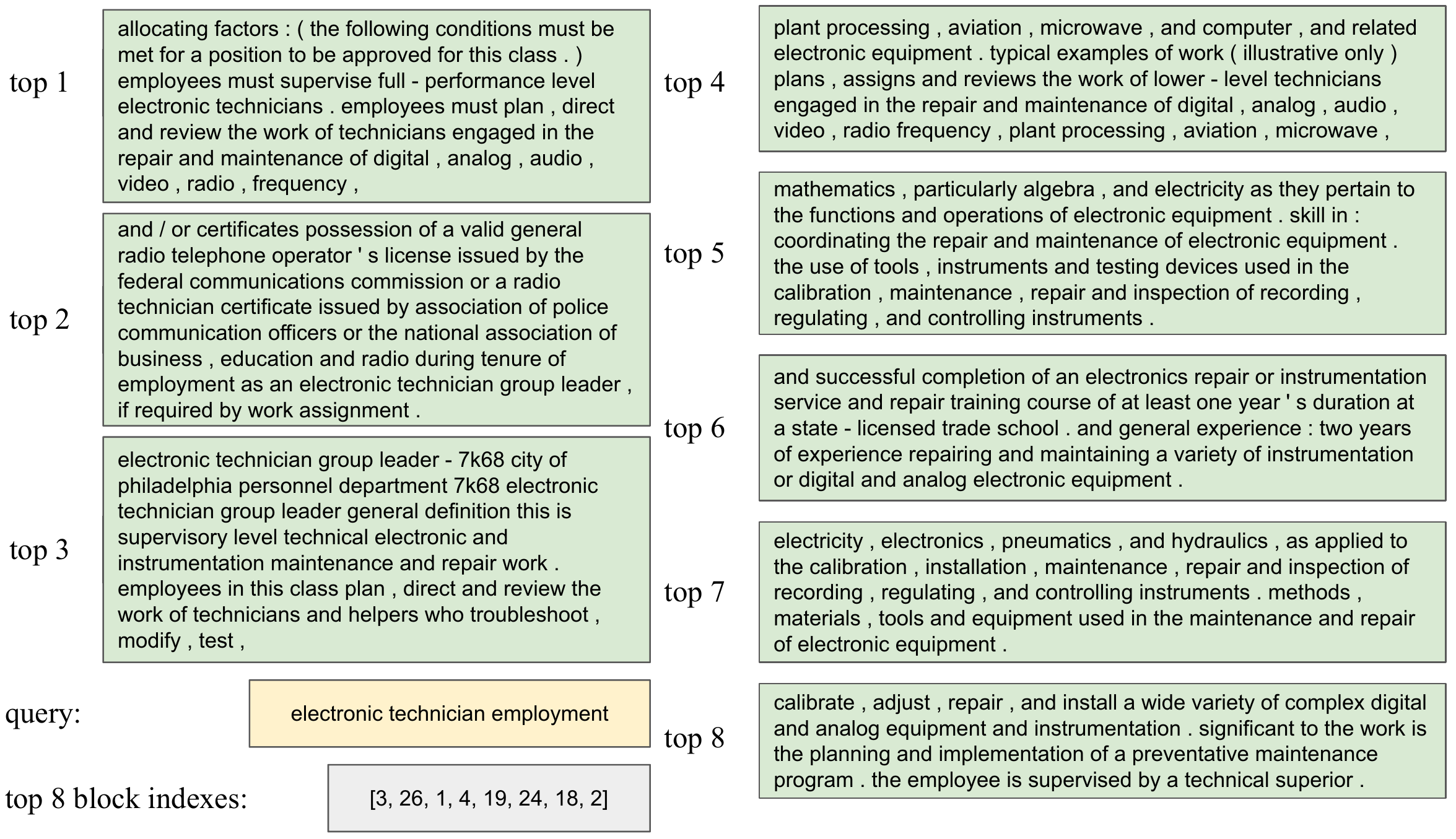}
  \caption{An example of top 8 blocks selected by the KeyB(vBERT)$_{BinB}$ model on MQ2007.}
  \label{fig.binbexample} 
  \end{figure}

To conclude this series of experiments on standard IR collections and for illustration purposes, we display an example in Figure~\ref{fig.binbexample} top 8 blocks selected by the KeyB(vBERT)$_{BinB}$ model on MQ2007. As one can see, the different blocks are distributed across different positions at the beginning, middle and end of the document. Furthermore, each block is related to the query.

\section{Experiment on TREC 2019 DL and comparison with sparse attention based models and IDCM}
\label{sec:experiments-2}

This section sheds additional light on the behaviour of the models proposed by comparing them to different baseline models, including sparse attention models and IDCM discussed in Section~\ref{sec:related-work}, on the relatively recent TREC DL dataset introduced to evaluate neural IR models. Table \ref{tab.dl19stat} summarizes the main characteristics of the the TREC 2019 Deep Learning Track collection. Following  \cite{jiang2020long}, we aim here to rerank the official top 100 retrieved documents and use the official evaluation metric NDCG@10 on the test set, which we complement with MAP. Since this collection is larger than previously used collections, we train the models, again using the pairwise hinge loss \cite{guo2016deep}, for longer steps, \textit{i.e.} for 10 epochs each epoch being composed of 15000 batches of 2 pairs (four documents). As the qrels for the training and validation sets only contain one annotated document for each query (and it is relevant), each pair is composed of a query, its relevant document and another randomly sampled document which is viewed as irrelevant. For each metric and each model, we select the hyper-parameters leading to the lowest loss on the validation set and report its performance on the test set. The other experimental settings are the same as those in Section~\ref{experDesignSec}.

\subsubsection{Comparison with sparse attention based models}
\label{Comparisonwithsparse}

Query-Directed Sparse Transformer (QDS-Transformer) \cite{jiang2020long} makes use of sparse local attention and global attention for long document information retrieval. In \cite{jiang2020long}, the experiments conducted on the TREC 2019 Deep Learning Track collection \cite{craswell2020overview} showed that QDS-Transformer improves the standard retrofitting BERT ranking baselines and outperforms more recent transformer architectures as Sparse Transformer \cite{child2019}, Longformer \cite{beltagy2020longformer}, and Transformer-XH \cite{Zhao2020Transformer-XH:}. We compare here our proposed approach with this QDS-Transformer and related baselines on this same collection.

\begin{table*}[tb]
    \centering
    \caption{Statistics of the TREC 2019 DL document ranking task.}
          \resizebox{\linewidth}{!}{
    \begin{tabular}{cccccccc} \toprule
    Collection & \# Documents & \# Train queries & \# Train qrels & \# Dev queries & \# Dev qrels & \# Test queries & \# Test qrels \\ \hline
      TREC19 DL & 3,213,835 & 367,013 & 384,597 & 5,193 & 5,478 & 43 & 16,258 \\
      \bottomrule
    \end{tabular}}
    \label{tab.dl19stat}
\end{table*}

\begin{table*}[t]
\centering
\caption{Experiment on TREC 2019 DL and comparison with sparse attention models and IDCM. Best results are in {\bf bold}.}
\label{tab:dl19}
\scalebox{1}{
\begin{tabular}{lll}
\toprule
\multicolumn{3}{r}{\textbf{TREC Deep Learning Track Document Ranking}} \\
\hline
Model   & NDCG@10 & MAP \\ 

\hline \hline \multicolumn{3}{l}{\textit{\textbf{Baseline models}}} \\
\hline
BM25 & 0.488 & 0.234 \\
CO-PACRR \cite{hui2017pacrr}       & 0.550 & 0.231 \\ 
TK \cite{DBLP:conf/ecai/HofstatterZH20}       & 0.594 & 0.252 \\ 
TKL \cite{hofstatter2020local}       & 0.644 & 0.277 \\ 
RoBERTa (FirstP) \cite{liu2019roberta, dai2019deeper}        & 0.588 & 0.233 \\ 
RoBERTa (MaxP) \cite{liu2019roberta, dai2019deeper}       & 0.630 & 0.246 \\ 
PARADE \cite{li2020parade}       & 0.655 & 0.280 \\
\hline \hline \multicolumn{3}{l}{\textit{\textbf{Sparse attention models}}} \\ \hline
Sparse-Transformer \cite{child2019}       & 0.634 & 0.257 \\ 
Longformer-QA \cite{beltagy2020longformer}     & 0.627 & 0.255 \\ 
Transformer-XH \cite{Zhao2020Transformer-XH:}      & 0.646 & 0.256 \\
QDS-Transformer \cite{jiang2020long}      & 0.667 & 0.278 \\
\hline \hline \multicolumn{3}{l}{\textit{\textbf{Select blocks models}}} \\ \hline
IDCM \cite{DBLP:conf/sigir/HofstatterMZCH21} & 0.679 & 0.273 \\
KeyB(vBERT)$_{BM25}$ & 0.678 & 0.277 \\
KeyB(vBERT)$_{BinB}$ & {\bf0.707} & {\bf0.281} \\
KeyB(PARADE5)$_{BM25}$ & 0.672 & 0.280 \\
KeyB(PARADE5)$_{BinB}$ & 0.676 & 0.277 \\
KeyB(PARADE5)$_{BinB2}$ & 0.678 & 0.279 \\
\bottomrule
\end{tabular}
}
\end{table*}

Table \ref{tab:dl19} shows the results obtained. 
Note that for PARADE, the number of passages is set to 16 and the max query length to 30 (other settings are the same as in Section~\ref{experDesignSec}). For the other models, we report the results given in \cite{jiang2020long}. As one can see, the best results are obtained with KeyB(vBERT)$_{BinB}$ which outperforms all baselines and sparse attention models, including QDS-Transformer, reaching 0.707 on NDCG@10 and 0.281 on MAP.  
It is closely followed by KeyB(vBERT)$_{BM25}$ which outperforms all baseline and sparse attention models on NDCG@10, reaching 0.678 compared with QDS-Transformer's 0.667. For MAP, KeyB(vBERT)$_{BM25}$ is slightly below QDS-Transformer and PARADE, and on a par with TKL. 

The KeyB(PARADE5) models are very close to the KeyB(vBERT) models, both in terms of NDCG@10 and MAP. They also outperform baseline and sparse attention models on NDCG@10, and outperform all baseline and sparse attention models but PARADE and QDS-Transformers on MAP (they are on a par with these to models on MAP). One can note however that on this collection the KeyB(PARADE5) models not as effective as on the previous collections NDCG@10. 
\citet{li2020parade} also observed this for the PARADE model and attributed it to the fact that the effectiveness of PARADE across collections is related to the number of relevant passages per document in these collections: TREC DL only has 1–2 relevant passages per document by construction; with such a low number of relevant passages, the benefit of utilizing complex passage aggregation methods such as PARADE is diminished. We see here however the advantage of the KeyB(PARADE5) models which rely on fewer d-passages, more likely to be relevant to the query. 

\subsubsection{Comparison with IDCM}

As mentioned in Section~\ref{sec:related-work}, IDCM \citep{DBLP:conf/sigir/HofstatterMZCH21} is a recently proposed model that also learns how to select blocks. The motivation behind this model is to obtain an IR model more efficient as it would only rely on a few blocks. Our motivation slightly differs as we aim to improve the overall IR system by filtering out non relevant, likely noisy blocks. Furthermore, our approach can be directly used with different IR models by selecting blocks with standard IR systems. This is the basis of the models KeyB(vBERT)$_{BM25}$ and KeyB(PARADE5)$_{BM25}$ for example.

Table~\ref{tab:dl19} shows a comparison of our approaches with IDCM (last four lines) where for each query the official top 100 documents are used (this setting is used for all models reported in Table~\ref{tab:dl19}). For IDCM, we have used the authors' notebook\footnote{\url{https://github.com/sebastian-hofstaetter/intra-document-cascade}}. IDCM reaches 0.679 for NDCG@10, which is higher than all baseline and sparse attention models, and 0.273 for MAP, which is higher than all baseline and sparse attention models but TKL, PARADE and QDS-Transformer. In contrast, KeyB(vBERT)$_{BinB}$ outperforms all models, including IDCM, on both evaluation metrics. In addition, both KeyB(vBERT)$_{BM25}$ and KeyB(PARADE5) variants, even though they did not benefit from an additional pre-training on MS-MARCO and rely on a much simpler procedure to select blocks, obtain results comparable with IDCM: KeyB(vBERT)$_{BM25}$ is $0.001$ point below IDCM on NDCG@10 and $0.004$ above on MAP whereas KeyB(PARADE5)$_{BM25}$ is $0.007$ point below IDCM on NDCG@10 and $0.007$ above on MAP, and KeyB(PARADE5)$_{BinB}$, KeyB(PARADE5)$_{BinB2}$ have higher MAP results.

Overall, the results obtained on TREC DL 2019 once again prove the effectiveness of the proposed selecting key blocks approaches, which obtain results than baseline and sparse attention models without the need to customize CUDA kernels. Besides, when selecting key blocks with the trained BERT model, which corresponds to the model KeyB(vBERT)$_{BinB}$, one obtains the near state-of-the-art level performance \cite{craswell2020overview} of 0.707 on this collection for NDCG@10. 
Lastly, the faster and less memory demanding variant KeyB(vBERT)$_{BM25}$ is a close competitor to KeyB(vBERT)$_{BinB}$ and should be preferred if time or memory constraints are important.

\section{Conclusion}
\label{sec:conclusion}

Benefiting from pre-trained BERT models, the field of information retrieval has seen remarkable progress in neural IR models, as exemplified by the success of Vanilla BERT which has become a strong, yet simple, baseline for neural IR models. To overcome the limitations of BERT-based models regarding long documents, we have proposed to divide documents into blocks and to select only the most important key blocks. This is reminiscent of the way humans assess the relevance of a document for a given query: one first identifies blocks relevant to the query, blocks which are then aggregated to obtain the overall assessment of the document. In order to select blocks, we have investigated two approaches: the first one is straightforward and makes use of standard retrieval functions as TF-IDF or BM25; the second one learns a single BERT model used for both ranking blocks and documents. Both approaches have been shown to improve over standard baselines and previous BERT-based models. We have followed the same approach on another highly competitive neural IR model, namely PARADE, here again with improved results. All in all, selecting blocks is advantageous for the two models studied here, Vanilla BERT and PARADE. We conjecture that this selection is a way to remove passages in documents which are not relevant to the query and which are likely to bring noise when matching queries and documents. 

Comparing our different proposals, if the selection strategy based on learned mechanisms performs in average better than the one based on standard IR models with similar GPU memory usage, its ranking latency is not as appealing. We thus recommend in practice to use the TF-IDF and BM25 versions of Vanilla BERT and PARADE if latency constraints are important (among these two variants, one may prefer the BM25 one which is slightly better overall). The choice between Vanilla BERT and PARADE variants, the latter being slightly better than the former on standard IR collections, and slightly worse on the TREC 2019 DL collection, depends on the collection considered.

In the future, we plan on deploying the proposed block selection approach on more complex models, which could be a way to further improve the results obtained in this study. We also plan to investigate alternative negative sampling strategies as well as ways to accelerate the selection process based on learned models.

\begin{acks}
  This work has been partially supported by MIAI@Grenoble Alpes (ANR-19-P3IA-0003) and the Chinese Scholarship Council (CSC) grant No.201906960018.
\end{acks}

\bibliographystyle{ACM-Reference-Format}
\bibliography{select-keyblocks}

\end{document}